\documentclass[11pt]{article}

\usepackage{amssymb}
\usepackage{epsfig}
\usepackage{graphics}
\usepackage{graphicx}
\usepackage{color}

\usepackage[latin1]{inputenc}
\usepackage[english]{babel}

\usepackage{amssymb}
\usepackage{amsmath}
\usepackage{amsfonts}
\numberwithin{equation}{section}

\input amssym.def
\input amssym.tex

\usepackage{booktabs}
\usepackage{ctable}
\usepackage{subfigure}
\usepackage{float}
\usepackage{layout}
\usepackage{fancyhdr}
\usepackage{color}
\usepackage{eurosym}
\usepackage{wrapfig}  

\usepackage{xr}

\def\ii{\mathbf{i}\,}

\def\I{\mbox{1}}
\def\P{{\mathbb P}}
\def\R{{\mathbb R}}
\def\E{{\mathbb E}}

\def\N{{\mathbb N}}

\def\bu{\overline{u}}
\def\bL{\overline{L}}

\def\cvd{$\quad\Box$\medskip}

\def\<{\langle}
\def\>{\rangle}

\def\Z{{\mathbb{Z}}}

\newtheorem{theorem}{Theorem}[section]

\newtheorem{lemma}[theorem]{Lemma}

\newtheorem{proposition}[theorem]{Proposition}
\newtheorem{remark}[theorem]{Remark}

\newcommand{\Dy}{\Delta y}

\newcommand{\ba}{\begin{eqnarray}}
\newcommand{\ea}{\end{eqnarray}}

\def \Sum{\displaystyle\sum}

\def \Frac{\displaystyle\frac}

\date{}

\begin{document}




\title{\bf 
Numerical stability of a hybrid method \\ 
for pricing options
}

\author{{\sc Maya Briani}\thanks{%
Istituto per le Applicazioni del Calcolo, CNR Roma - {\tt m.briani@iac.cnr.it}}\\
{\sc Lucia Caramellino}\thanks{%
Dipartimento di Matematica,
Universit\`a di Roma Tor Vergata, and INdAM-GNAMPA - {\tt caramell@mat.uniroma2.it}}\\
{\sc Giulia Terenzi}\thanks{%
Dipartimento di Matematica,
Universit\`a di Roma Tor Vergata, and LAMA, Universit\'e Paris-Est - {\tt terenzi@mat.uniroma2.it}}\\
{\sc Antonino Zanette}\thanks{%
Dipartimento di Scienze
Economiche e Statistiche,
Universit\`a di Udine - {\tt antonino.zanette@uniud.it}}}
%
\maketitle
\begin{abstract}\noindent{\parindent0pt
We develop and study stability properties of a hybrid approximation of functionals of the Bates jump model with stochastic interest rate that uses a tree method in the direction of the volatility and the interest rate and a finite-difference approach in order to handle the underlying asset price process. We also propose hybrid simulations for the model, following a binomial tree in the direction of both the volatility and the interest rate, and a space-continuous approximation for the underlying asset price process coming from a Euler-Maruyama type scheme. We test our numerical schemes by computing  European and American option prices.
}
\end{abstract}

\noindent \textit{Keywords:}  stochastic volatility; jump-diffusion process; European and American options; tree methods; finite-difference; numerical stability.

\smallskip

\noindent \textit{2000 MSC:} 91G10, 60H30, 65C20.


%
%
%
%
%
%
%

\section{Introduction}


Following the work in \cite{bcz,bcz-hhw},  we further develop and study the hybrid tree/finite-difference approach and the hybrid Monte Carlo technique in order to numerically evaluate option prices. We concern here with  the  theoretical study of the stability of the  numerical scheme for both European and American options. Also, in this paper we stress the model (and the associated numerical procedure) by considering the Bates model \cite{bates}, possibly coupled with a stochastic interest rate following the Vasicek dynamics \cite{Vas}, and we call the full model as Bates-Hull-White.

The option pricing tree/finite-difference approach 
we deal with,  derives from applying an efficient simple (recombining binomial) 
tree method in the direction of the volatility and the interest rate
components, whereas the asset price component is locally treated by
means of a one-dimensional partial integro-differential equation
(PIDE), to which a finite-difference scheme is applied. Here, the
numerical treatment of the nonlocal term coming from the jumps
involves implicit-explicit techniques, as well as numerical
quadratures. The procedure applies to other rather general stochastic volatility models (see \cite{bct}) or for problems in insurance (see \cite{GMZ}).
We concentrate here our attention to the Bates-Hull-White model and we give results on the numerical stability.
Let us mention that, to this purpouse, we never require the validity of the Feller condition for the
Cox-Ingersol-Ross (CIR) dynamics \cite{cir} of the volatility
process.

In the case of plain vanilla European options, Fourier inversion methods \cite{CM} lead to closed-form formulas to compute the price under the Bates model. For American options, numerical  methods are typically based on the use of the dynamic programming principle to which one applies either deterministic schemes for solutions of PIDEs from numerical analysis and/or from tree methods or Monte Carlo techniques. Our method is a mixture between a  tree approach and a deterministic numerical approach, and is particularly tailored for the use of the backward dynamic programming principle.

Let us recall that tree methods for the Heston model have been already proposed in the literature.  For example,
Akyildirim, Dolinsky and Soner  \cite{ads} have recently provided  a four tuple discrete Markov approximation which can be generalized to other stochastic volatility models with a factor equation. Lo, Nguyen and Skindilias \cite{lnd} have proposed a Markov chain approximation based on a trinomial tree, again adaptable to other stochastic volatility models.  Vellekoop and Nieuwenhuis \cite{VN} have introduced a binomial tree built from the full truncation scheme of Lord, Koekkoek and Dijk \cite{lkd}. Other tree approaches for the Heston model are available (see e.g. the references quoted in \cite{VN}). Generally, the Feller condition for the CIR volatility equation is required, either for theoretical purposes or for the numerical efficiency of the method. 

Another tool is given by the dicretization of PDEs.
When the jumps are not considered, 
available references are recalled in \cite{bcz, bcz-hhw}. In the standard Bates model, 
the finite-difference methods for solving the $2$-dimensional PIDE associated with the option pricing problems can be based on implicit, explicit or alternating direction implicit schemes. The implicit scheme requires to solve a dense sparse system at each time step. Toivanen \cite{t} proposes a componentwise splitting method for pricing American options. 
In Ballestra and Cecere \cite{bct}, the problem is handled 
by using an ad hoc pseudospectral method. 
Chiarella, Kang, Meyer and Ziogas \cite{ckmz}
developed a method of lines algorithm for pricing and hedging American
options again under the standard Bates dynamics.
Itkin \cite{it}  has recently proposed a unified approach to handle PIDEs associated with L\'evy's models of interest in Finance.

From the simulation point of view, the main problem consists in the treatment of the CIR dynamics for the volatility process. 
Several efficient and accurate methods have been  developed specifically for the simulation of CIR paths, see, e.g.,  Alfonsi \cite{al}, Andersen \cite{andersen}, Lord, Koekkoek and Dijk \cite{lkd} or Kahl and J\"{a}ckel \cite{kj}. 
We propose here a hybrid Monte Carlo technique: we couple the simulation of the approximating tree for the volatility and the interest rate components with a standard simulation of the underlying asset price, based on Brownian increments and a straightforward treatment of the jumps. In the case of American option, this is associated with the Longstaff and Schwartz algorithm \cite{ls}, allowing to treat the dynamic programming principle. The numerical results are then compared with the Alfonsi's  third-order simulation scheme.

%

The paper is organized as follows. In Section \ref{sect-model}, we
introduce the Bates-Hull-White model. In Section \ref{discretization} we recall the tree procedure for the volatility and the interest rate pair (Section \ref{sect-tree}), we describe our discretization of the log-price process (Section \ref{sec:approxY}) and  the hybrid Monte Carlo simulations (Section \ref{sect-MC}). Section \ref{sect:htfd} is devoted to the hybrid tree/finite-difference method: we set the numerical scheme for  the associated local PIDE problem (Section \ref{sec:approxPDE}) and we apply it to the solution of the whole pricing scheme (Section \ref{sect-alg}). Section \ref{sect:stability} is devoted to the analysis of the numerical stability of the resulting tree/finite-difference method.
Section \ref{practice} refers to the practical use of our methods. Here, numerical results and comparisons are widely discussed.

\section{The Bates-Hull-White model}\label{sect-model}

The Bates model \cite{bates} is a stochastic volatility model
with price jumps: the dynamics of the underlying asset price is driven
by both a Heston stochastic volatility \cite{hes} and a compound Poisson jump process of the type originally introduced by Merton \cite{mer}.
We allow the interest rate to follow a stochastic model, which we assume to be a generalized Ornstein-Uhlenbeck (hereafter OU) process.
More precisely, the dynamics under the risk neutral measure of the share price $S$, the volatility process $V$ and the interest rate $r$, are given by the following jump-diffusion model:
\begin{equation}\label{BHHmodel}
\begin{array}{l}
\displaystyle\frac{dS_t}{S_{t^-}}= (r_t-\eta)dt+\sqrt{V_t}\, dZ^S_t+d H_t,
\smallskip\\
dV_t= \kappa_V(\theta_V-V_t)dt+\sigma_V\sqrt{V_t}\,dZ^V_t,
\smallskip\\
dr_t= \kappa_r(\theta_r(t)-r_t)dt+\sigma_r dZ^r_t,
\end{array}
\end{equation}
where $\eta$ denotes the continuous dividend rate, $S_0,V_0,r_0>0$, $Z^S$, $Z^V$ and $Z^r$ are correlated Brownian motions and $H$ is a compound Poisson process with
intensity $\lambda$ and i.i.d. jumps $\{J_k\}_k$, that is
\begin{equation}\label{H}
H_t=\sum_{k=1}^{K_t} J_k,
\end{equation}
$K$ denoting a Poisson process with intensity $\lambda$.
We assume that the Poisson process $K$, the jump amplitudes $\{J_k\}_k$ and the $3$-dimensional correlated Brownian motion $(Z^S,Z^V,Z^r)$ are independent. As suggested by Grzelak and Oosterlee in \cite{GO}, the significant correlations are between the noises governing the pairs $(S,V)$ and $(S,r)$. So, as done in \cite{bcz-hhw}, we assume that the couple $(Z^V,Z^r)$ is a standard Brownian motion in $\R^2$ and $Z^S$ is a Brownian motion in $\R$ which is correlated both with $Z^V$ and $Z^r$:
$$
d\<Z^S,Z^V\>_t=\rho_1dt \ \mbox{ and }\ d\<Z^S,Z^r\>_t=\rho_2dt.
$$
We recall that the volatility process $V$ follows a CIR dynamics with mean reversion rate $\kappa_V$, long run variance $\theta_V$ and $\sigma_V$ denotes the vol-vol (volatility of the volatility). We assume that $\theta_V,\kappa_V,\sigma_V>0$ and we stress that we never require in this paper that the CIR process satisfies the Feller condition $2\kappa_V\theta_V\geq \sigma_V^2$, ensuring that the process $V$ never hits $0$. So, we allow the volatility $V$ to reach $0$. The interest rate $r_t$ is described by a generalized OU process, in particular $\theta_r$ is time-dependent but deterministic and fits the zero-coupon bond market values, for details see  \cite{bm}. As already done in \cite{hw1}, we write the process $r$ as follows:
\begin{equation}\label{rX}
r_t   = \sigma_r X_t + \varphi_t
\end{equation}
where
\begin{equation}\label{Xphi}
X_t  = - \kappa_r \int_0^tX_s\,ds + \,Z^r_t \quad\mbox{and}\quad
\varphi_t=r_0e^{-\kappa_r t}+\kappa_r\int_0^t\theta_r(s)e^{-\kappa_r(t-s)}ds.
\end{equation}


\medskip

From now on we set
$$
Z^V=W^1,\quad Z^r=W^2, \quad Z^S=\rho_1 W^1+\rho_2 W^2+ \rho_3 W^3,
$$
where $W=(W^1,W^2,W^3)$ is a standard Brownian motion in $\R^3$ and the correlation parameter $\rho_3$ is given by
$$
\rho_3=\sqrt{1-\rho_1^2-\rho_2^2},\quad \rho_1^2+\rho_2^2\leq 1.
$$
By passing to the logarithm $Y=\ln S$ in the first component, by taking into account the above mentioned correlations and by considering the process $X$ as in \eqref{rX}-\eqref{Xphi}, we get the triple $(Y,V,X)$ given by
\begin{equation}\label{YVX-dyn}
\begin{array}{ll}
&dY_t=\mu_Y(V_t,X_t, t)dt+\sqrt{V_t}\, \big(\rho_1dW^1_t+\rho_2dW^2_t+\rho_3dW^3_t\big) + dN_t, \quad Y_0=\ln S_0\in\R, \smallskip\\
&dV_t= \mu_V(V_t)dt+\sigma_V\sqrt{V_t}\,dW^1_t,
\quad V_0>0,\smallskip\\
&dX_t=\mu_X(X_t)dt+dW^2_t,
\quad X_0=0,
\end{array}
\end{equation}
where
\begin{align}
\label{muY}
&\mu_Y(v,x,t)=\sigma_rx+\varphi_t-\eta-\frac 12 \,v,\\
\label{muV}
&\mu_V(v)=\kappa_V(\theta_V-v),\\
\label{muX}
&\mu_X(x)= -\kappa_r x,
\end{align}
and $N_t$ is the compound Poisson process written through the Poisson process $K$, with intensity $\lambda$, and the i.i.d. jumps $\{\log(1+J_k)\}_k$, that is
$$
N_t=\sum_{k=1}^{K_t} \log (1+J_k),
$$
Recall that $K$, the jump amplitudes $\{\log(1+J_k)\}_k$ and the $3$-dimensional standard Brownian motion $(W^1,W^2, W^3)$ are all independent.
We also recall that the L\'evy measure associated with $N$ is given by
$$
\nu(dx)=\lambda \P(\log(1+J_1)\in dx),
$$
and whenever $\log (1+J_1)$ is absolutely continuous then $\nu$ has a density as well:
\begin{equation}\label{nu}
\nu(dx)=\nu(x)dx=\lambda p_{\log(1+J_1)}(x)dx,
\end{equation}
$p_{\log(1+J_1)}$ denoting the probability density function of $\log(1+J_1)$.
For example, in the Merton model \cite{mer} it is assumed that $\log(1+J_1)$ has a normal distribution - this is the choice we will do in our numerical experiments,
as done in Chiarella \textit{et al.} \cite{ckmz}. But other jump-amplitude measures can be selected. For instance, in the Kou model \cite{kou} the law of $\log(1+J_1)$ is a mixture of exponential  laws:
$$
p_{\log(1+J_1)}(x) = p\lambda_+ e^{-\lambda_+ x}\,\I_{\{x>0\}}+(1-p)\lambda_- e^{\lambda_- x}\,\I_{\{x<0\}},
$$
$\I_A$ denoting the indicator function of $A$.  Here, the parameters $\lambda_\pm>0$ control the decrease of the distribution tails of negative and positive jumps respectively, and $p$ is the  probability of a positive jump.

We pass to the transformation $Y=\ln S$. If $\Psi(Y)$ denotes the payoff written on the log-price,
the option price $P=P(t,y,v,x)$ is given by
\begin{equation}\label{price}
\begin{array}{ll}
\mbox{European price: }&
\displaystyle
P(t,y,v,x)
=\E\Big(e^{-\int_t^T (\sigma_rX^{t,x}_s+\varphi_s)ds}\Psi(Y^{t,y,v,x}_T)\Big),\smallskip\\
\mbox{American price: } &
\displaystyle
P(t,y,v,x)=\sup_{\tau\in \mathcal{T}_{t,T}}\E\Big(e^{-\int_t^\tau (\sigma_rX^{t,x}_s+\varphi_s)ds}\Psi(Y^{t,y,v,x}_\tau)\Big),
\end{array}
\end{equation}
where $\mathcal{T}_{t,T}$ denotes the set of all stopping times taking values on $[t,T]$. 
Hereafter,  $(Y^{t,y,v,x},V^{t,v},X^{t,x})$ denotes the solution of the jump-diffusion dynamic \eqref{YVX-dyn} starting at time $t$ in the point $(y,v,x)$.


\section{The dicretized process}\label{discretization}

We first set up the discretization of the triple $(Y,V,X)$ we will take into account.

\subsection{The 2-dimensional tree for $(V,X)$}\label{sect-tree}

We consider an approximation for the pair $(V,X)$ in \eqref{YVX-dyn} on the time-interval $[0,T]$ by means of a $2$-dimensional computationally simple tree. This means that we construct a Markov chain running over a $2$-dimensional recombining bivariate lattice and, at each time-step, both components of the Markov chain can jump only upwards or downwards. We consider the ``multiple-jumps'' approach by Nelson and  Ramaswamy \cite{nr}, extensively developed for the CIR process in \cite{acz}.
We give here the main ideas  in order to set-up the whole algorithm. We start by considering a discretization of the time-interval $[0,T]$ in $N$ subintervals $[nh,(n+1)h]$, $n=0,1,\ldots,N$, with $h=T/N$.

For $n=0,1,\ldots,N$, consider the lattice for $V$ and $X$ defined by
\begin{align}
&\mathcal{V}_n =\{v^n_k\}_{k=0,1,\ldots,n}\quad\mbox{with}\quad
v^n_k=\Big(\sqrt {V_0}+\frac{\sigma_V} 2(2k-n)\sqrt{h}\Big)^2\I_{\{\sqrt {V_0}+\frac{\sigma_V} 2(2k-n)\sqrt{h}>0\}}, \label{state-space-V}
\\
&\mathcal{X}_n =\{x^n_j\}_{j=0,1,\ldots,n}\quad\mbox{with}\quad
x^n_j=(2j-n)\sqrt{h}, \label{state-space-X}
\end{align}
respectively.
Notice that $v_{0,0}=V_0$ and $x_{0,0}=0=X_0$. For each fixed $v^n_k\in\mathcal{V}_n $ and $x^n_j\in \mathcal{X}_n$, we denote the ``up'' and ``down'' jump by $v^{n+1}_{ k_u (n,k)}$ and $v^{n+1}_{k_d (n,k)}$ and by $x^{n+1}_{ j_u (n,j)}$ and $x^{n+1}_{j_d (n,j)}$. By applying the ``multiple jump approach'', the jump-indexes  $k_u (n,k)$, $k_d (n,k)$, $j_u (n,j)$, $j_d (n,j)$ are defined  as
\begin{align}
\label{ku}
&k_u (n,k) =\min\{k^*\,:\, k+1\leq k^*\leq n+1\mbox{ and }v^n_k+\mu_V(v^n_k)h \le v^{n+1}_{k^*}\},\\
\label{kd}
&k_d (n,k) =\max\{k^*\,:\, 0\leq k^*\leq k \mbox{ and }v^n_k+\mu_V(v^n_k)h \ge v_{n+1, k^*}\},\\
\label{ju}
&j_u (n,j) =\min\{j^*\,:\, j+1\leq j^*\leq n+1\mbox{ and }x^n_j+\mu_X(x^n_j)h \le x^{n+1}_{j^*}\},\\
\label{jd}
&j_d (n,j) =\max\{j^*\,:\, 0\leq j^*\leq j \mbox{ and }x^n_j+\mu_X(x^n_j)h \ge x^{n+1}_{j^*}\},
\end{align}
where  $\mu_V$ and $\mu_X$ is is the drift of $V$ and $X$ respectively (see \eqref{muV} and \eqref{muX}), 
with the understanding $k_u (n,k)=n+1$, respectively $k_d (n,k)=0$, if the set in the r.h.s. of \eqref{ku}, respectively  \eqref{kd}, is empty, and similarly for $j_u$ and $j_d$. 


\begin{figure}[htp]
	\centering
	\begin{tabular}{c}
		
		\includegraphics[scale=0.5]{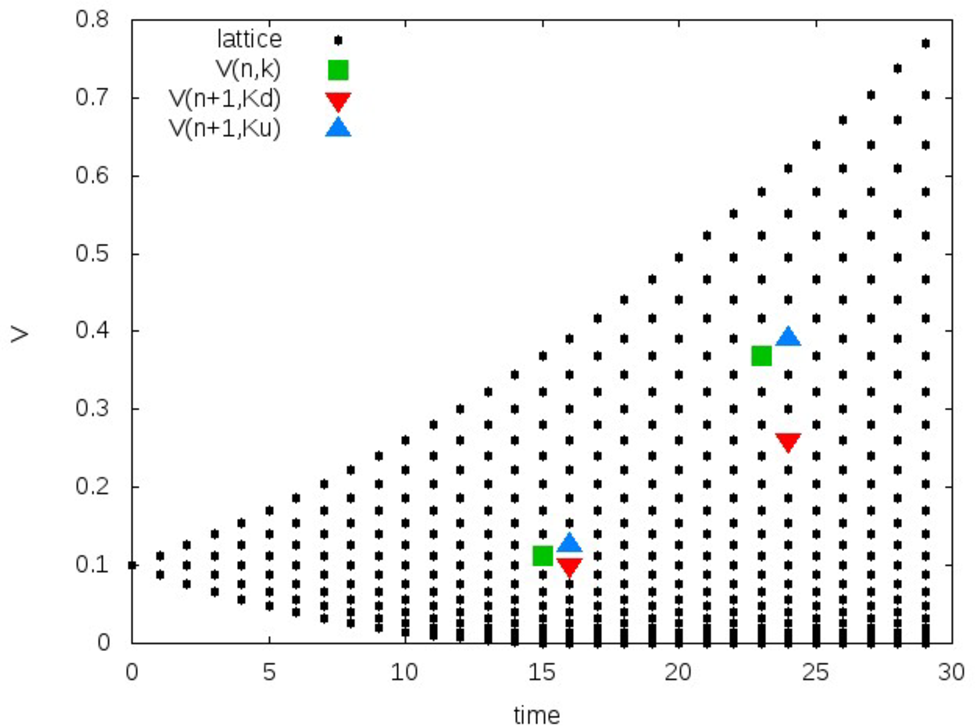}
		\includegraphics[scale=0.5]{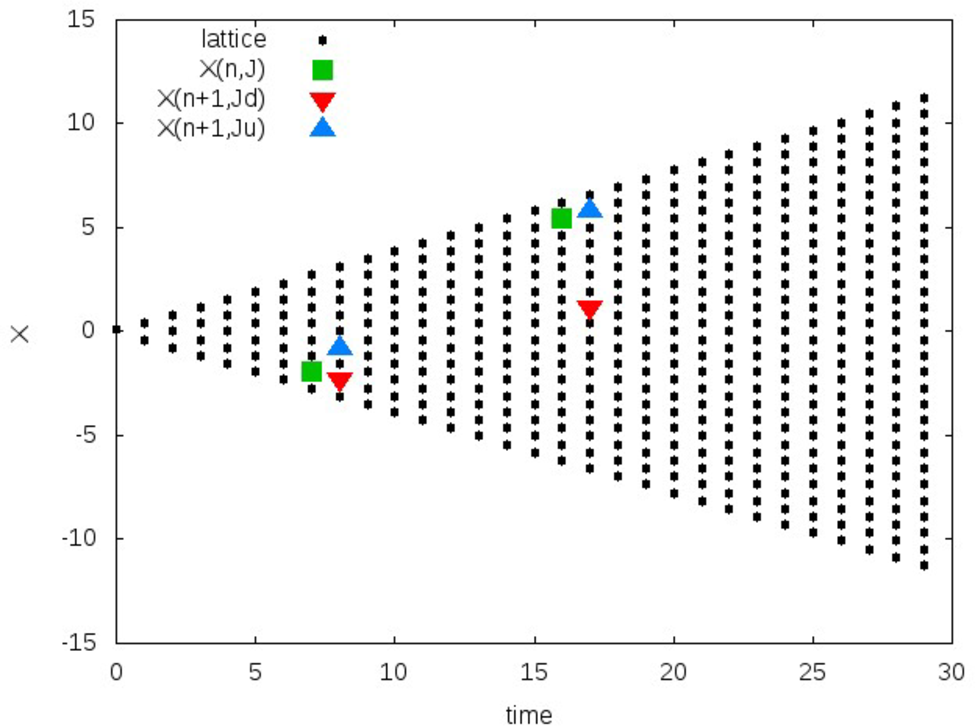}
		
	\end{tabular}
	\caption{\em\small The tree for the process $V$ (left) and for $X$ (right). The figure shows possible instances of the up and down jumps. }
	\label{fig:treeVX}
\end{figure}

The transition probabilities for $V$ are defined as follows: starting from the node $(n,k)$ the probability that the process  jumps to $k_u (n,k)$ and $k_d (n,k)$ at time-step $n+1$ are set as
\begin{equation}\label{pik}
p^{V}_u(n,k)
=0\vee \frac{\mu_V(v^n_k)h+ v^n_k-v^{n+1}_{k_d (n,k)} }{v^{n+1}_{k_u (n,k)}-v^{n+1}_{k_d (n,k)}}\wedge 1
\quad\mbox{and}\quad p^{V}_d(n,k)=1-p^{V}_u(n,k)
\end{equation}
respectively. A similar definition is set for the component $X$: starting from $(n,j)$, the probability that the process jumps to $j_u (n,j)$ and $j_d (n,j)$ at time-step $n+1$ are 
\begin{equation}\label{pij}
p^{X}_u(n,j)
=0\vee \frac{\mu_X(x^n_j)h+ x^n_j-x^{n+1}_{j_d (n,j)} }{x^{n+1}_{j_u (n,j)}-x^{n+1}_{j_d (n,j)}}\wedge 1
\quad\mbox{and}\quad p^{X}_d(n,j)=1-p^{X}_u(n,j)
\end{equation}
respectively.

%
%
%


%
%
%
%

\smallskip

The tree procedure for the pair $(V,X)$  is obtained by joining the trees built for $V$ and for $X$. Namely, for $n=0,1,\ldots,N$, consider the lattice
\begin{equation}\label{state-space-Vr}
\mathcal{V}_n \times \mathcal{X}_n =\{(v^n_k,x^n_j)\}_{k,j=0,1,\ldots,n}.
\end{equation}
Starting from the node $(n,k,j)$, which corresponds to the position $(v^n_k,x^n_j)\in\mathcal{V}_n \times \mathcal{X}_n $, we define the four possible jumps by means of the following four nodes at time $n+1$:
\begin{equation}\label{treescheme}
\begin{array}{lcr}
(n+1,k_u (n,k),j_u (n,j)) & \hbox{with probability} & p_{uu}(n,k,j)=p^{V}_{u}(n,k)p^{X}_{u}(n,j), \smallskip\\
(n+1,k_u (n,k),j_d (n,j)) & \hbox{with probability} & p_{ud}(n,k,j)=p^{V}_{u}(n,k)p^{X}_{d}(n,j), \smallskip\\
(n+1,k_d (n,k),j_u (n,j)) & \hbox{with probability} & p_{du}(n,k,j)=p^{V}_{d}(n,k)p^{X}_{u}(n,j), \smallskip\\
(n+1,k_d (n,k),j_d (n,j)) & \hbox{with probability} & p_{dd}(n,k,j)=p^{V}_{d}(n,k)p^{X}_{d}(n,j),
\end{array}
\end{equation}
where the above nodes $k_{u}(n,k)$, $k_{d}(n,k)$, $j_{u}(n,j)$, $j_{d}(n,j)$ and
the above probabilities $p^{V}_{u}(n,k)$, $p^{V}_{d}(n,k)$, $p^{X}_{u}(n,j)$, $p^{X}_{d}(n,j)$ are defined in \eqref{ku}-\eqref{kd}, \eqref{ju}-\eqref{jd}, \eqref{pik} and \eqref{pij}. The factorization of the jump probabilities in \eqref{treescheme} follows from the orthogonality property of the noises driving the two processes.
This procedure gives rise to a Markov chain $(\hat V^h_{n},\hat X^h_{n})_{n=0,\ldots,N}$ that weakly converges on the path space, as $h\to 0$, to the diffusion process $(V_{t},X_t)_{t\in[0,T]}$ solution to
\begin{align*}
&dV_t= \mu_V(V_t)dt+\sigma_V\sqrt{V_t}\,dW^1_t,
\quad V_0>0,\\
&dX_t  = \mu_X(X_t)\,dt + dW^2_t,\quad X_0=0.
\end{align*}
This can be proved by using standard results (see e.g. the techniques in \cite{nr}) and the convergence of the chain approximating the volatility process proved in \cite{bib:acz}. And  this holds independently of the validity of the Feller condition $2\kappa_V \theta_V\geq \sigma^2_V$.

Details and remarks on the extension of this procedure to more general cases can be found in \cite{bcz-hhw}. In particular, if the correlation between the Brownian motions driving $(V,X)$ was not null, one could define the jump probabilities by matching the local cross-moment (see Remark 3.1 in \cite{bcz-hhw}).

\subsection{The approximation of the component $Y$}\label{sec:approxY}

We describe here how we manage the $Y$-component in \eqref{YVX-dyn} by taking into account the tree procedure given for the pair $(V,X)$.
We go back to \eqref{YVX-dyn}: by isolating $\sqrt{V_t}dW^1_t$ in the second line and $dW^2_t$ in the third one, we obtain
\begin{equation}\label{Y1}
dY_t=\mu(V_t,X_t,t)dt+\rho_3\sqrt{V_t}\,dW^3_t+\frac{\rho_1}{\sigma_V}dV_t+\rho_2\sqrt{V_t}dX_t+dN_t \end{equation}
with
\begin{equation}\label{mu-fin}
\begin{array}{rl}
\mu(v,x,t)
&=\mu_Y(v,x,t)-\frac{\rho_1}{\sigma_V}\mu_V(v)-\rho_2\sqrt{v}\,\mu_X(x)\smallskip\\
&=\sigma_rx+\varphi_t-\eta-\frac 12 \,v
-\frac{\rho_1}{\sigma_V}\,\kappa_V(\theta_V-v)+\rho_2\kappa_r x\sqrt v
\end{array}
\end{equation}
($\mu_Y$, $\mu_V$ and $\mu_X$ are defined in \eqref{muY}, \eqref{muV} and \eqref{muX} respectively). To numerically solve \eqref{Y1}, we mainly use the fact that the noises $W^3$ and $N$ are independent of the processes $V$ and $X$.
So, we first take the approximating tree $(\hat V^h_n,\hat X_n)_{n=0,1,\ldots,N-1}$ discussed in Section \ref{sect-tree} and we set $(\bar V^h_t,\bar X^h_t)_{t\in[0,T]}=(\hat V^h_{\lfloor t/h\rfloor},\hat X^h_{\lfloor t/h\rfloor})_{t\in[0,T]}$ the associated time-continuous c\`adl\`ag approximating process for $(V,X)$.
Then, we insert the discretization $(\bar V^h,\bar X^h)$ for $(V,X)$ in the coefficients of \eqref{Y1}. Therefore, the final process $\bar Y^h$  approximating $Y$ is set as follows:
$\bar Y^h_0=Y_0$ and for $t\in(nh,(n+1)h]$ with $n=0,1,\ldots,N-1$
\begin{equation}\label{barYh}
\begin{array}{rl}
\bar Y^h_t=& \bar Y^h_{nh} + \mu(\bar V^h_{nh},\bar X^h_{nh},nh)(t-nh)
+\rho_3\sqrt{\bar V^h_t}(W^3_t-W^3_{nh})
\medskip\\
&+\displaystyle\frac{\rho_1}{\sigma_V}(\bar V^h_t-\bar V^h_{nh})+\rho_2\sqrt{\bar V^h_t}(\bar X^h_t-\bar X^h_{nh})+(N_t-N_{nh}).
\end{array}
\end{equation}

\subsection{The Monte Carlo approach}\label{sect-MC}
Let us show how one can simulate a single path by using the tree approximation \eqref{state-space-Vr} for the couple $(V,X)$ and the Euler scheme \eqref{barYh} for the $Y$-component.

Let $(\hat Y_n)_{n=0,1,\ldots,N}$ be the sequence approximating $Y$ at times $nh$, $n=0,1,\ldots,N$,  by means of the scheme in \eqref{barYh}: $\hat Y^h_0=Y_0$ and for $t\in[nh,(n+1)h]$ with $n=0,1,\ldots,N-1$ then
$$
\begin{array}{rl}
\hat Y^h_{n+1}=& \hat Y^h_{n} + \mu(\hat V^h_{n},\hat X^h_{n},nh)h
+\rho_3\sqrt{h\hat V^h_n}\Delta_{n+1}
\medskip\\
&+\displaystyle\frac{\rho_1}{\sigma_V}(\hat V^h_{n+1}-\hat V^h_{n})+\rho_2\sqrt{\hat V^h_n}(\hat X^h_{n+1}-\hat X^h_{n})+(N_{(n+1)h}-N_{nh}),
\end{array}
$$
where $\mu$ is defined in \eqref{mu-fin} and $\Delta_1,\ldots,\Delta_N$ denote i.i.d. standard normal r.v.'s, independent of the noise driving the chain $(\hat V,\hat X)$. The simulation of $N_{(n+1)h}-N_{nh}$ is straightforward: one first generates a Poisson r.v. $K_h^{n+1}$ of parameter $\lambda h$ and if $K_h^{n+1}>0$ then also the log-amplitudes $\log(1+J^{n+1}_k)$ for $k=1,\ldots,K^{n+1}_h$ are simulated. Then, the observed jump of the compound Poisson process is written as the sum of the simulated log-amplitudes, so that
\begin{equation}\label{hmc}
\begin{array}{rl}
\hat Y^h_{n+1}=& \hat Y^h_{n} + \mu(\hat V^h_{n},\hat X^h_{n},nh)h
+\rho_3\sqrt{h\hat V^h_n}\Delta_{n+1}
\medskip\\
&+\displaystyle\frac{\rho_1}{\sigma_V}(\hat V^h_{n+1}-\hat V^h_{n})+\rho_2\sqrt{\hat V^h_n}(\hat X^h_{n+1}-\hat X^h_{n})+\sum_{k=1}^{K_h^{n+1}}\log(1+J^{n+1}_k),
\end{array}
\end{equation}
in which the last sum is set equal to 0 if $K_h^{n+1}=0$.

The above simulation scheme is plain: at each time step $n\geq 1$, one let the pair $(V,X)$ evolve on the tree and simulate the process $Y$ by using \eqref{hmc}. We will refer to this procedure as \textit{hybrid Monte Carlo algorithm}, the word ``hybrid'' being related to the fact that two different noise sources are considered: we simulate a continuous process in space (the component $Y$) starting from a discrete process in space (the tree for $(V,X)$).

The simulations just described will be used in next Section \ref{practice} in order to set-up a Monte Carlo procedure for the computation of the option price function  \eqref{price}. In the case of American options, the simulations are coupled with the Monte Carlo algorithm by Longstaff and Schwartz in \cite{ls}.

\section{The hybrid tree/finite difference approach}\label{sect:htfd}

The option price in \eqref{price} is typically computed by  means of the standard backward dynamic programming algorithm. So,  consider a discretization of the time interval $[0,T]$ into $N$ subintervals of length $h=T/N$. Then the price $P(0,Y_0,V_0,X_0)$  is numerically approximated through the quantity $P_h(0,Y_0,V_0,X_0)$ backwardly given by
\begin{equation}\label{backward}
  \begin{cases}
    P_h(T,y,v,x)= \Psi(y)\quad
     \mbox{and as } n=N-1,\ldots,0,\\
   P_h(nh,y,v,x) =  \max \Big\{\widehat \Psi(y), e^{-(\sigma_rx+\varphi_{nh})h}
   \E\Big(P_h\big((n+1)h, Y_{(n+1)h}^{nh,y,v,x}, V_{(n+1)h}^{nh,v}, X_{(n+1)h}^{nh,x}\big)
   \Big)\Big\},
  \end{cases}
\end{equation}
for $(y,v,x)\in\R\times\R_+\times \R$, in which
$$
\widehat \Psi(y)=\left\{
\begin{array}{ll}
0 & \mbox{ in the European case,}\smallskip\\
\Psi(y) & \mbox{ in the American case.}
\end{array}
\right.
$$
So, what is needed is a good approximation of the expectations appearing in the above dynamic programming principle. This is what we first deal with, starting from the dicretized process $(\bar Y^h,\bar V^h,\bar X^h)$ introduced in Section \ref{discretization}.

\subsection{The local 1-dimensional partial integro-differential equation}\label{sec:approxPDE}
Let $\bar Y^h$ denote the process in \eqref{barYh}.
If we set
\begin{equation}\label{Z}
\bar Z^{h}_t=\bar Y^h_t
-\frac{\rho_1}{\sigma_V}(\bar V^h_t-\bar V^h_{nh})
-\rho_2\sqrt{\bar V^h_{nh}}(\bar X^h_t-\bar X_{nh}), \quad t\in [nh,(n+1)h]
\end{equation}
then we have ($\mu$ being given in \eqref{mu-fin})
\begin{equation}\label{barZ}
\begin{array}{l}
d\bar Z^{h}_t=\mu(\bar V^h_{nh},\bar X^h_{nh},{nh})dt+\rho_3\sqrt{\bar V^h_{nh}}\,dW^3_t,+dN_t\quad t\in (nh,(n+1)h],\smallskip\\
\bar Z^{h}_{nh}=\bar Y^h_{nh},
\end{array}
\end{equation}
that is, $\bar Z^h$ solves a jump-diffusion stochastic equation with constant coefficients and at time $nh$ it starts from $\bar Y^h_{nh}$.
Take now a function $f$: we are interested in computing
$$
\E(f(Y_{(n+1)h})\mid Y_{nh}=y, V_{nh}=v, X_{nh}=x).
$$
We actually need a function $f$ of the whole variables $(y,v,x)$ but at the present moment the variable $y$ is the most important one, we will see later on that the introduction of $(v,x)$ is straightforward. So, we numerically compute the above expectation by means of the one done on the approximating processes, that is,
$$
\begin{array}{l}
\displaystyle
\E\big(f(\bar Y^h_{(n+1)h})\mid \bar Y^h_{nh}=y, \bar V^h_{nh}=v, \bar X^h_{nh}=x\big)\smallskip\\
\displaystyle
\qquad=\E\big(f(\bar Z^{h}_{(n+1)h}+\frac{\rho_1}{\sigma_V}(\bar V^h_{(n+1)h}-\bar V^h_{nh})
+\rho_2\sqrt{\bar V^h_{nh}}(\bar X^h_{(n+1)h}-\bar X^h_{nh})
)\mid \bar Z^{h}_{nh}=y, \bar V^h_{nh}=v, \bar X^h_{nh}=x\big),
\end{array}
$$
in which we have used the process $\bar Z^h$ in \eqref{Z}. Since $(\bar V^h, \bar X^h)$ is independent of the Brownian noise $W^3$ and on the compound Poisson process $N$ driving $\bar Z^h$ in \eqref{barZ}, we have the following:
we set
\begin{equation}\label{E2}
\Psi_f(\zeta; y,v,x)=\E(f(\bar Z^h_{(n+1)h}+\zeta
)\mid \bar Z^h_{nh}=y, \bar V^h_{nh}=v, \bar X^h_{nh}=x)
\end{equation}
and we can write
\begin{equation}\label{E1}
\begin{array}{l}
\displaystyle
\E(f(\bar Y^h_{(n+1)h})\mid \bar Y^h_{nh}=y, \bar V^h_{nh}=v, \bar X^h_{nh}=x)\smallskip\\
\qquad=\E\Big(\Psi_f\Big(\frac{\rho_1}{\sigma_V}(\bar V^h_{(n+1)h}-\bar V^h_{nh})
+\rho_2\sqrt{v}(\bar X^h_{(n+1)h}-\bar X^h_{nh}); y,v,x
\Big)\,\Big|\, \bar V^h_{nh}=v, \bar X^h_{nh}=x\Big).
\end{array}
\end{equation}
Now, in order to compute the quantity $\Psi_f(\zeta)$ in \eqref{E2}, we consider a generic function $g$ and set
$$
u(t,y;v,x)=\E(g(\bar Z^h_{(n+1)h})\mid \bar Z^h_{t}=y, \bar V^h_{t}=v, \bar X^h_{t}=x),\quad t\in[nh,(n+1)h].
$$
By \eqref{barZ} and the Feynman-Kac representation formula we can state that, for every fixed $x\in\R$ and $v\geq 0$, the function $(t,y)\mapsto u(t,y;v,x)$ is the solution to
\begin{equation}\label{PIDE}
\left\{\begin{array}{ll}
\displaystyle
\partial_t u(t,y;v,x)+L^{(v,x)} u(t,y;v,x)=0 & y\in\R, t\in [nh,(n+1)h),\smallskip\\
\displaystyle
u((n+1)h,y;v,x)=g(y) & y\in \R,
\end{array}\right.
\end{equation}
where $L^{(v,x)}$ is the integro-differential operator
\begin{equation}\label{PIDE:Loperator}
\begin{array}{rl}
L^{(v,x)}u(t,y;v,x)=&\mu(v,x)\partial_y u(t,y;v,x) +\frac 12 \rho_3^2 v\partial^2_{yy} u(t,y;v,x)
\medskip\\
&+ \displaystyle\int_{-\infty}^{+\infty} \left[ u(t,y+\xi;v,x)-u(t,y;v,x)\right]\nu(\xi) d\xi,
\end{array}
\end{equation}
where $\mu$ is given in \eqref{mu-fin} and $\nu$ is the L\'evy measure associated with the compound Poisson process $N$, see \eqref{nu}. We are assuming here that the L\'evy measure is absolutely continuous (in practice, we use a Gaussian density), but it is clear that the procedure we are going to describe can be straightforwardly extended to other cases.

\subsubsection{Finite-difference and numerical quadrature}

In order to numerically compute the solution to the PIDE \eqref{PIDE} at time $nh$,
we generalize the approach already developed in \cite{bcz,bcz-hhw}:
we apply a one-step finite-difference algorithm to the differential part of the problem coupled now with a quadrature rule to approximate the integral term.

We start by fixing an infinite grid on the $y$-axis $\mathcal{Y}=\{y_i=Y_0+i\Dy\}_{i\in\Z}$, with $\Delta y=y_i-y_{i-1}$, $i\in\Z$. For fixed $n$ and given $x\in\R$ and $v\geq 0$, we set $u^n_i=u(nh, y_i;v,x)$ the discrete solution of \eqref{PIDE} at time $nh$ on the point $y_i$ of the grid $\mathcal{Y}$ -- for simplicity of notations, in the sequel we do not stress in $u^n_i$ the dependence on $(v,x)$.

First of all, to numerically compute the integral term in \eqref{PIDE:Loperator} we need to truncate the infinite integral domain to a bounded interval $\mathcal{I}$, to be taken large enough in order that
\begin{equation}\label{bound_int}
\int_{\mathcal{I}} \nu(\xi) d\xi \approx \lambda.
\end{equation}
In terms of the process, this corresponds to truncate the large jumps.
We assume that the tails of $\nu$ rapidly decrease -- this is not really restrictive since applied  models typically require that the tails of $\nu$ decrease exponentially.
Hence, we take $R\in\N$ large enough, set $\mathcal{I}=[-R\Delta y , +R\Delta y ]$ and  apply to \eqref{bound_int} the trapezoidal rule on the grid $\mathcal{Y}$ with the same step $\Dy$ previously defined.  Then, for  $\xi_l=l\Dy$, $l=-R,\ldots,R$, we have
\begin{equation}\label{num_int}
\int_{-R\Delta y}^{+R\Delta y} \left[ u(t,y+\xi)-u(t,y)\right]\nu(\xi) d\xi\approx \Dy \sum_{l=-R}^{R} \left(u(t,y+\xi_l)-u(t,y)\right)\nu(\xi_l).
\end{equation}
We notice that $y_i+\xi_l=Y_0+(i+l)\Dy \in\mathcal{Y}$, so the values $u(t,y_i+\xi_l)$ are well defined on the numerical grid $\mathcal{Y}$ for any $ i,l$. These are technical settings and can be modified and calibrated for different L\'evy measures $\nu$.

But in practice one cannot solve the PIDE problem over the whole real line. So, we have to choose artificial bounds and impose numerical boundary conditions. We take a positive integer $M>0$ and we define a finite grid $\mathcal{Y}_M=\{y_i=Y_0+i\Dy\}_{i\in\mathcal{J}_{M}}$, with $\mathcal{J}_{M}=\{-M,\ldots,M\}$, and we assume that $M>R$.
Notice that 
for $y=y_i\in \mathcal{Y}_M$ then the integral term in \eqref{num_int} splits into two parts: one part concerning nodes falling into the numerical domain $\mathcal{Y}_{M}$ and another part concerning nodes falling out of $\mathcal{Y}_M$. As an example, at time $t=nh$ we have
$$
\sum_{l=-R}^{R}u(nh,y_i+\xi_l)\nu(\xi_l)\approx\sum_{l=-R}^R u^{n}_{i+l} \nu(\xi_l) = \sum_{l\,:\,|l|\leq R,|i+l| \leq M} u^{n}_{i+l}\  \nu(\xi_l) + \sum_{l\,:\,|l|\leq R,|i+l| > M} \tilde u^{n}_{i+l}\  \nu(\xi_l),
$$
where $\tilde u^n_{\cdot}$ stands for (unknown) values that fall out of the finite numerical domain $\mathcal{Y}_{M}$.
This implies that we must choose some suitable artificial boundary conditions.
In a financial context, in \cite{cv} it has been shown that a good choice for the boundary conditions is the payoff function.
Although this is the choice we will do in our numerical experiments, for the sake of generality we assume here the boundary values outside $\mathcal{Y}_{M}$ to be settled as $\tilde u^{n}_{i}=b(nh,y_i)$, where $b=b(t,y)$ is a fixed function defined in
$[0,T]\times\R$.

Going back to the numerical scheme to solve the differential part of the equation \eqref{PIDE}, as already done in \cite{bcz-hhw}, we apply an implicit in time approximation. However, to avoid to solve at each time step a linear system with a dense matrix, the non-local integral term needs anyway an explicit in time approximation. We then obtain an implicit-explicit (hereafter IMEX) scheme as proposed in \cite{cv} and \cite{bln}. Notice that more sophisticated IMEX methods may be applied, see for instance \cite{bnr,st}. Let us stress that these techniques could be used in our framework, being more accurate but expensive.

As done in \cite{bcz-hhw}, to achieve greater precision we use the centered approximation for both first and second order derivatives in space.
The discrete solution $u^n$ at time $nh$ is then computed in terms of the known value $u^{n+1}$ at time $(n+1)h$ by solving the following discrete problem: for all $i\in\mathcal{J}_M$,
\begin{equation}\label{discr_eq}
\Frac{u^{n+1}_i-u^n_i}{h}+\tilde\mu_Y(v,x)\Frac{u^{n}_{i+1}-u^{n}_{i-1}}{2\Dy}+\frac{1}{2}\rho_3^2\ v\ \Frac{u^{n}_{i+1}-2u^n_i+u^{n}_{i-1}}{\Dy^2}
 + \Dy \Sum_{l=-R}^{R} \left(u^{n+1}_{i+l}-u^{n+1}_{i}\right)\nu(\xi_l) =0.
\end{equation}
We then get the solution $u^n=(u^n_{-M},\ldots,u^n_M)^T$ by solving the following linear system
\begin{equation}\label{lin_sys}
A\, u^n = B u^{n+1} + d,
\end{equation}
where $A=A(v,x)$ and $B$ are $(2M+1)\times (2M+1)$ matrices and $d$ is a $(2M+1)$-dimensional boundary vector defined as follows.

\smallskip

\noindent
\textbf{1. The matrix $A$.}
From \eqref{discr_eq}, we set $A$ as the tridiagonal real matrix given by
\begin{equation}\label{matrixA}
A = \left(
\begin{array}{ccccc}
1+2\beta & -\alpha-\beta & & & \\
\alpha-\beta & 1+2\beta & -\alpha-\beta & & \\
 & \ddots & \ddots & \ddots &  \\
 &   & \alpha-\beta & 1+2\beta & -\alpha-\beta \\
 &   &        &   \alpha-\beta & 1+2\beta
\end{array}
\right),
\end{equation}
with
\begin{equation}\label{alphabeta_impl}
\alpha=\frac{h}{2\Dy}\,\mu(nh,v,x)\quad\mbox{and}\quad\beta=\frac{h}{2\Dy^2}\,\rho_3^2v,
\end{equation}
$\mu$ being defined in \eqref{mu-fin}.
We stress on that at each time step $n$, the quantities $v$ and $x$ are constant and known values (defined by the tree procedure for $(V,X)$) and then $\alpha$ and $\beta$ are constant parameters.

\smallskip

\noindent
\textbf{2. The matrix $B$.}
Again from \eqref{discr_eq}, $B$ is the $(2M+1)\times(2M+1)$ real matrix given by
\begin{equation}\label{matrixD_bound}
B = I + h\Dy\left(
\begin{array}{cccccc}
 \nu(0) - \Lambda &  \nu(\Dy)  & \ldots & \nu(R\Dy) & 0 & \\
\nu(-\Dy) & \nu(0) - \Lambda & \nu(\Dy)  & \ldots &  \nu(R\Dy) & \\
 & \ddots & \ddots & \ddots &  &\\
0 & \nu(-R\Dy) & \ldots & \nu(-\Dy) &  \nu(0) - \Lambda
\end{array}
\right),
\end{equation}
where $I$ is the identity matrix and
$$
\Lambda=\sum_{l=-R}^R \nu(\xi_l).
$$

\medskip

\noindent
\textbf{3. The boundary vector $d$.}
The vector $d\in\R^{2M+1}$ contains the numerical boundary values:
\begin{equation}\label{d}
d = a_b^n+ a_b^{n+1},
\end{equation}
with
$$a_b^n =((\beta-\alpha)b^{n}_{-M-1},0,\ldots,0,(\beta+\alpha)b^n_{M+1})^T\in\R^{2M+1}$$
and $a_b^{n+1} \in \R^{2M+1}$ is such that
$$
(a^{n+1}_b)_i = \left\{\begin{array}{ll}
\displaystyle
h \Dy \sum_{l=-R}^{-M-i-1}\nu(x_l)\ b^{n+1}_{i+l}, & \mbox{ for } i=-M,\ldots,-M+R-1,
\smallskip\\
0 & \mbox{ for } i=-M+R,\ldots,M-R,
\smallskip\\
\displaystyle
h \Dy \sum_{l=M-i+1}^R\nu(x_l)\ b^{n+1}_{i+l}, & \mbox{ for } i=M-R+1,\ldots,M-1,
\end{array}\right.
$$
where we have used the standard notation $b^n_i=b(nh,y_i)$, $i\in\mathcal{J}_M$.

\smallskip
In practice, we numerically solve the linear system \eqref{lin_sys} with an efficient algorithm (see next Remark \ref{costo}). We notice here that a solution to \eqref{lin_sys} really exists
because for $\beta\ne|\alpha|$, the matrix $A=A(v,x)$ is invertible (see e.g. Theorem 2.1 in \cite{bt}). Then, at time $nh$, for each fixed $v\geq 0$ and $x\in\R$,
we approximate the solution $y\mapsto u(nh,y ;v,x)$ of \eqref{PIDE} on the points $y_i$'s of the grid in terms of the discrete solution $u^n=\{u^n_i\}_{i\in \mathcal{J}_{M}}$, which in turn is written in terms of the value $u^{n+1}=\{u^{n+1}_i\}_{i\in \mathcal{J}_{M}}$ at time $(n+1)h$. In other words, we set
\begin{equation}\label{FD}
\mbox{$u(nh,y_i;v,x)\approx u^n_i$, $i\in\mathcal{J}_M$, where $u^n=(u^n_i)_{i\in\mathcal{J}_M}$ solves \eqref{lin_sys}}
\end{equation}


\subsubsection{The final local finite-difference approximation}

We are now ready to tackle our original problem: the computation of the function $\Psi_f(\zeta;y,v,x)$ in \eqref{E2} allowing one to numerically compute the expectation in \eqref{E1}.
So, at time step $n$, the pair $(v,x)$ is chosen on the lattice $\mathcal{V}_n\times\mathcal{X}_n$: $v=v^n_k$, $x=x^n_j$ for $0\leq k,j\leq n$. We call $A^n_{k,j}$ the matrix $A$ in \eqref{matrixA} when evaluated in  $(v^n_k, x^n_j)$ and $d^n$ the boundary vector in \eqref{d} at time-step $n$. Then, \eqref{FD} gives
$$
\begin{array}{c}
\mbox{$\Psi_f(\zeta; y_i,v^n_k,x^n_j)\simeq u^n_{i,k,j}$, where $u^n_{\cdot,k,j}=(u^n_{i,k,j})_{i\in\mathcal{J}_M}$ solves the linear system}\medskip\\
\displaystyle
A^n_{k,j}u^n_{\cdot,k,j}=Bf(y_\cdot+\zeta)+d^n.
\end{array}
$$
Therefore, by taking the expectation w.r.t. the tree-jumps, the expectation in \eqref{E1} is finally computed on
$\mathcal{Y}_M\times\mathcal{V}_n\times\mathcal{X}_n$ by means of the above approximation:
$$
\begin{array}{l}
\mbox{$\E(f(\bar Y^h_{(n+1)h})\mid \bar Y^h_{nh}=y_i, \bar V^h_{nh}=v^n_k, \bar X^h_{nh}=x^n_j)\simeq u^n_{i,k,j}$,}\smallskip\\
\mbox{where $u^n_{\cdot,k,j}=(u^n_{i,k,j})_{i\in\mathcal{J}_M}$ solves the linear system}\smallskip\\
\displaystyle
A^n_{k,j}u^n_{\cdot,k,j}\smallskip\\
\displaystyle
\quad=\sum_{a,b\in\{u,d\}} p_{ab}(n,k,j) Bf
\Big(y_\cdot+\frac{\rho_1}{\sigma_V}(v^{n+1}_{k_a (n,k)}-v^n_k)
+\rho_2\sqrt{v}(x^{n+1}_{j_b (n,j)}-x^n_j)\Big)+
d^n.
\end{array}
$$
Finally, if $f$ is a function on the whole triple $(y,v,x)$, by using standard properties of the conditional expectation one gets
\begin{equation}\label{E3}
\begin{array}{l}
\mbox{$\E(f(\bar Y^h_{(n+1)h}, \bar V^h_{(n+1)h}, \bar X^h_{(n+1)h})\mid \bar Y^h_{nh}=y_i, \bar V^h_{nh}=v^n_k, \bar X^h_{nh}=x^n_j)\simeq u^n_{i,k,j}$,}\smallskip\\
\mbox{where $u^n_{\cdot,k,j}=(u^n_{i,k,j})_{i\in\mathcal{J}_M}$ solves the linear system}\smallskip\\
\displaystyle
A^n_{k,j}u^n_{\cdot,k,j}\smallskip\\
\displaystyle
=\sum_{a,b\in\{u,d\}} p_{ab}(n,k,j) Bf
\Big(y_\cdot+\frac{\rho_1}{\sigma_V}(v^{n+1}_{k_a (n,k)}-v^n_k)
+\rho_2\sqrt{v}(x^{n+1}_{j_b (n,j)}-x^n_j), v^{n+1}_{k_a (n,k)}, x^{n+1}_{j_b (n,j)}\Big)+
d^n.
\end{array}
\end{equation}

\subsection{Pricing European and American options}\label{sect-alg}

We are now ready to approximate the function $P_h$ solution to the dynamic programming principle \eqref{backward}. We consider the discretization scheme $(\bar Y^h,\bar V^h,\bar X^h)$ discussed in Section \ref{sec:approxPDE} and we use the approximation \eqref{E3} for the conditional expectations that have to be computed at each time step $n$. So, for every point $(y_i,v^n_k, x^n_j)\in \mathcal{Y}_{M}\times\mathcal{V}_n\times\mathcal{X}_n$, by \eqref{E3} we have
$$
\E\Big(P_h\big((n+1)h, Y_{(n+1)h}^{nh,y_{i},v^n_k,x^n_j}, V_{(n+1)h}^{nh,v^n_k}, X_{(n+1)h}^{nh,x^n_j}\big)\Big)\simeq u^n_{i,k,j}
$$
where $u^n_{\cdot,k,j}=(u^n_{i,k,j})_{i\in\mathcal{J}_M}$ solves the linear system
\begin{equation}\label{E5}
\begin{array}{l}
\displaystyle
A^n_{k,j}u^n_{\cdot,k,j}=
B\!\!\!\sum_{a,b\in\{u,d\}}\!\!\! p_{ab}(n,k,j)\times \smallskip\\
\displaystyle
\times P_h
\Big((n+1)h,y_\cdot+\frac{\rho_1}{\sigma_V}(v^{n+1}_{k_a (n,k)}-v^n_k)
+\rho_2\sqrt{v}(x^{n+1}_{j_b (n,j)}-x^n_j, v^n_k,x^n_j), v^{n+1}_{k_a (n,k)}, x^{n+1}_{j_b (n,j)}\Big)+d^n.
\end{array}
\end{equation}
We then define the approximated price $\tilde P_h(nh,y,v,x)$ for $(y,v,x)\in \mathcal{Y}_{M}\times\mathcal{V}_n\times\mathcal{X}_n$ and $n=0,1,\ldots,N$ as
\begin{equation}\label{backward-ter0}
\begin{cases}
\tilde P_h(T,y_i,v_{N,k},x_{N,j})= \Psi(y_i)
\quad \mbox{and as $n=N-1,\ldots,0$:}\\
\displaystyle
\tilde P_h(nh,y_i,v^n_k,x^n_j) =
\max \Big\{\widehat \Psi(y_i),
e^{-(\sigma_r x^n_j + \varphi_{nh})h}
\tilde u^n_{i,k,j}\Big\}
  \end{cases}
\end{equation}
in which $\tilde u^n_{\cdot,k,j}=(\tilde u^n_{i,k,j})_{i\in\mathcal{J}_M}$ is the solution to the system in \eqref{E5} with $P_h$ replaced by $\tilde P_h$.

Note that the system in \eqref{E5} requires the knowledge of the function $y\mapsto \tilde P_h((n+1)h,y,v,x)$ in points $y$'s that do not necessarily belong to the grid $\mathcal{Y}_M$. Therefore, in practice we compute such a function by means of linear  interpolations, working as follows.
For fixed $n,k,j,a,b$, we set $I_{n,k,j,a,b}(i)$, $i\in\mathcal{J}_M$, as the index such that
$$
y_i+\frac{\rho_1}{\sigma_V}(v^{n+1}_{k_a (n,k)}-v^n_k)
+\rho_2\sqrt{v^n_k}(x^{n+1}_{j_b (n,j)}-x^n_j)\in[y_{I_{n,k,j,a,b}(i)}, y_{I_{n,k,j,a,b}(i)+1}),
$$
with $I_{n,k,j,a,b}(i)=-M$ if $y_i+\frac{\rho_1}{\sigma_V}(v^{n+1}_{k_a (n,k)}-v^n_k)
+\rho_2\sqrt{v^n_k}(x^{n+1}_{j_b (n,j)}-x^n_j)<-M$ and
$I_{n,k,j,a,b}(i)+1=M$ if $y_i+\frac{\rho_1}{\sigma_V}(v^{n+1}_{k_a (n,k)}-v^n_k)
+\rho_2\sqrt{v^n_k}(x^{n+1}_{j_b (n,j)}-x^n_j)>M$. We set
$$
q_{n,k,j,a,b}(i)
=\frac{y_i+\frac{\rho_1}{\sigma_V}(v^{n+1}_{k_a (n,k)}-v^n_k)
+\rho_2\sqrt{v^n_k}(x^{n+1}_{j_b (n,j)}-x^n_j)-y_{I_{n,k,j,a,b}(i)}}{\Delta y}.
$$
Note that $q_{n,k,j,a,b}(i)\in[0,1)$. We  define
\begin{align*}
&(\mathfrak{I}_{a,b}\tilde P_h )((n+1)h,y_i, v^{n+1}_{k_a (n,k)}, x^{n+1}_{j_b (n,j)})
=\tilde P_h ((n+1)h,y_{I_{n,k,j,a,b}(i)}, v^{n+1}_{k_a (n,k)}, x^{n+1}_{j_b (n,j)})\,(1-q_{n,k,j,a,b}(i))\\
&\quad +\tilde P_h ((n+1)h,y_{I_{n,k,j,a,b}(i)+1}, v^{n+1}_{k_a (n,k)}, x^{n+1}_{j_b (n,j)})\,q_{n,k,j,a,b}(i)
\end{align*}
and we set
$$
\begin{array}{l}
\displaystyle
\tilde P_h\Big((n+1)h,y_i+\frac{\rho_1}{\sigma_V}(v^{n+1}_{k_a (n,k)}-v^n_k)
+\rho_2\sqrt{v}(x^{n+1}_{j_b (n,j)}-x^n_j), v^{n+1}_{k_a (n,k)}, x^{n+1}_{j_b (n,j)}\Big)\smallskip\\
=(\mathfrak{I}_{a,b}\tilde P_h)((n+1)h,y_i, v^{n+1}_{k_a (n,k)}, x^{n+1}_{j_b (n,j)}).
\end{array}
$$
Therefore, starting from \eqref{E5}, in practice the function $\tilde u^n_{\cdot,k,j}=(\tilde u^n_{i,k,j})_{i\in\mathcal{J}_M}$ in \eqref{backward-ter0} is taken as the solution to the linear system
\begin{equation}\label{u-interp}
A^n_{k,j}\tilde u^n_{\cdot,k,j}
= B\sum_{a,b\in\{u,d\}}\!\!\! p_{ab}(n,k,j)
(\mathfrak{I}_{a,b}\tilde P_h)((n+1)h,y_\cdot, v^{n+1}_{k_a (n,k)}, x^{n+1}_{j_b (n,j)})+ d^n.
\end{equation}
We can then state our final numerical procedure:
\begin{equation}\label{backward-ter}
\begin{cases}
\tilde P_h(T,y_i,v_{N,k},x_{N,j})= \Psi(y_i)\quad 
\mbox{and as $n=N-1,\ldots,0$:}\\
\displaystyle
\tilde P_h(nh,y_i,v^n_k,x^n_j) =
\max \Big\{\widehat \Psi(y_i),
e^{-(\sigma_r x^n_j + \varphi_{nh})h}
\tilde u^n_{i,k,j}\Big\}
  \end{cases}
\end{equation}
$\tilde u^n_{\cdot,k,j}=(\tilde u^n_{i,k,j})_{i\in\mathcal{J}_M}$ being the solution to the system \eqref{u-interp}.

\begin{remark}\label{interp}
In the case of infinite grid, that is $M=+\infty$, $i\mapsto I_{n,k,j,a,b}(i)$ is a translation: $I_{n,k,j,a,b}(i)=I_{n,k,j,a,b}(0)+i$. So, $y_i\mapsto (\mathfrak{I}_{a,b}\tilde P_h )((n+1)h,y_i, v^{n+1}_{k_a (n,k)}, x^{n+1}_{j_b (n,j)})$ is just a linear convex combination of a translation of $y_i\mapsto \tilde P_h ((n+1)h,y_i, v^{n+1}_{k_a (n,k)}, x^{n+1}_{j_b (n,j)})$.
\end{remark}

\section{Stability analysis of the hybrid tree/finite-difference method}\label{sect:stability}

For the study of the stability, we consider a norm on the functions of  $(y,v,x)$ which is uniform with respect to the volatility and the interest rate components $(v,x)$ and coincides with the standard $l_2$ norm with respect to the direction $y$ (see next \eqref{norm}). The choice of the $l_2$ norm allows one to perform a von Neumann analysis  in the component $y$ on the infinite grid $\mathcal{Y}=\{y_i=Y_0+i\Dy\}_{i\in\Z}$, that is, without truncating the domain and without imposing boundary conditions. Therefore, our stability analysis does not take into account boundary effects. This approach is extensively used in the literature, see e.g. \cite{duffy}, and yields good criteria on the robustness of the algorithm independently of the boundary conditions.

Let us first write down explicitly the scheme \eqref{backward-ter} on the infinite grid $\mathcal{Y}=\{y_i\}_{i\in\Z}$.
For a fixed  function $f=f(t,y,v,x)$, we set $g=f$ either $g=0$ and we consider the numerical scheme given by
 \begin{equation} \label{backward-ter-inf-noL}
	\begin{cases}
		F_h(T,y_i,v_{N,k},x_{N,j})= f(T,y_i,v_{N,k},x_{N,j})\quad
		\mbox{and as $n=N-1,\ldots,0$:}\\
		\displaystyle
		F_h(nh,y_i,v^n_k,x^n_j) =\max \Big\{g(nh,y_i,v^n_k,x^n_j),
		e^{-(\sigma_r x^n_j + \varphi_{nh})h} u^n_{i,k,j}\Big\}
	\end{cases}
\end{equation}
where $u^n_{\cdot,k,j}=(u^n_{i,k,j})_{i\in\Z}$ is the solution to
	\begin{equation}\label{scheme_bates}
	\begin{array}{l} (\alpha_{n,k,j}-\beta_{n,k})u^n_{i-1,k,j}+(1+2\beta_{n,k})u^n_{i,k,j}-(\alpha_{n,k,j}+\beta_{n,k})u^n_{i+1,k,j}
	\smallskip\\
\displaystyle
	=\sum_{a,b\in\{d,u\}}p_{ab}(n,k,j)\times\Big[ (\mathfrak{I}_{a,b}F_h)((n+1)h,y_i, v^{n+1}_{k_a (n,k)}, x^{n+1}_{j_b (n,j)})+\smallskip\\
\displaystyle
+h\Dy\sum_l\nu(\xi_l)\big((\mathfrak{I}_{a,b}F_h)((n+1)h, y_{i+l}, v^{n+1}_{k_a (n,k)}, x^{n+1}_{j_b (n,j)})
-(\mathfrak{I}_{a,b}F_h)((n+1)h, y_{i}, v^{n+1}_{k_a (n,k)}, x^{n+1}_{j_b (n,j)})\big)\Big],
	\end{array}
	\end{equation}
in which $\alpha_{n,k,j}$ and $\beta_{n,k,j}$ are the coefficients $\alpha$ and $\beta$ defined in \eqref{alphabeta_impl} when evaluated in the pair $(v^n_k, x^n_j)$. Note that \eqref{scheme_bates} is simply the  linear system \eqref{u-interp} on the infinite grid, with $d^n\equiv 0$ (no boundary conditions are needed). Let us stress that in next Remark \ref{antitrasf} we will see that, since $\beta_{n,k}\geq 0$, a solution to \eqref{scheme_bates} does exist, at least for ``nice'' functions $f$. It is clear that the case $g=f$ is linked to the American algorithm whereas the case $g=0$ is connected to the European one: \eqref{backward-ter-inf-noL} gives our numerical approximation of the function
\begin{equation}\label{F}
F(t,y,v,x)=\left\{
\begin{array}{ll}
\displaystyle
\E\Big(e^{-(\sigma_r\int_t^TX^{t,x}_sds+\int_t^T\varphi_sds)}f(T,Y_T^{t,y,v,x},V^{t,v}_T, X^{t,x}_T)\Big)
&\mbox{ if } g=0,\smallskip\\
\displaystyle
\sup_{\tau\in\mathcal{T}_{t,T}}\E\Big(e^{-(\sigma_r\int_t^{\tau}X_s^{t,x}ds+\int_t^\tau \varphi_sds)}f(\tau,Y_\tau^{t,y,v,x},V^{t,v}_\tau, X^{t,x}_\tau)\Big)
&\mbox{ if } g=f,
\end{array}
\right.
\end{equation}
at times $nh$ and in the points of the grid $\mathcal{Y}\times \mathcal{V}_n\times \mathcal{X}_n$.

\subsection{The ``discount truncated scheme'' and its stability}

In our stability analysis, we consider a numerical scheme which is a slightly modification of \eqref{backward-ter-inf-noL}: we fix a (possibly large) threshold $L>0$ and we consider the scheme
 \begin{equation} \label{backward-ter-inf}
	\begin{cases}
		F^L_h(T,y_i,v_{N,k},x_{N,j})= f(T,y_i,v_{N,k},x_{N,j})\quad 
		\mbox{and as $n=N-1,\ldots,0$:}\\
		\displaystyle
		F^L_h(nh,y_i,v^n_k,x^n_j)
=\max \Big\{g(nh,y_i,v^n_k,x^n_j),
		e^{-(\sigma_r x^n_j\,1_{\{x^n_j>-L\}} + \varphi_{nh})h}
        u^n_{i,k,j}\Big\}
	\end{cases}
\end{equation}
with $g=f$ or $g=0$, where $u^n_{\cdot,k,j}=(u^n_{i,k,j})_{i\in\Z}$ is the solution to \eqref{scheme_bates}, with $(\mathfrak{I}_{a,b}F_h)$ replaced by
$(\mathfrak{I}_{a,b}F^L_h)$.
Let us stress that the above scheme \eqref{backward-ter-inf}  really differs from  \eqref{backward-ter-inf-noL} only when $\sigma_r>0$ (stochastic interest rate). And in this case, in the discounting factor of \eqref{backward-ter-inf} we do not allow  $x^n_j$ to run everywhere on its grid: in the original scheme \eqref{backward-ter-inf-noL}, the exponential contains the term $x^n_j$ whereas in the present scheme \eqref{backward-ter-inf} we put $x^n_j1_{\{x^n_j>-L\}}$, so we kill the points of the grid $\mathcal{X}_n$ below the threshold $-L$. And in fact, \eqref{backward-ter-inf} aims to numerically compute the function
\begin{equation}\label{FL}
F^L(t,y,v,x)=\left\{
\begin{array}{ll}
\displaystyle
\E\Big(e^{-(\sigma_r\int_t^TX^{t,x}_s\,1_{\{X^{t,x}_s>-L\}}ds
+\int_t^T\varphi_sds)}f(T,Y_T^{t,y,v,x},V^{t,v}_T, X^{t,x}_T)\Big)
&\mbox{ if } g=0,\smallskip\\
\displaystyle
\sup_{\tau\in\mathcal{T}_{t,T}}
\E\Big(e^{-(\sigma_r\int_t^{\tau}X_s^{t,x}\,1_{\{X^{t,x}_s>-L\}}ds+\int_t^\tau \varphi_sds)}f(\tau,Y_\tau^{t,y,v,x},V^{t,v}_\tau, X^{t,x}_\tau)\Big)
&\mbox{ if } g=f,
\end{array}
\right.
\end{equation}
at times $nh$ and in the points of the grid $\mathcal{Y}\times \mathcal{V}_n\times \mathcal{X}_n$.
Recall that in practice $h$ is small but fixed, therefore  there is a natural threshold which actually comes on in practice (see for instance the tree given in Figure \ref{fig:treeVX}). And actually, in our numerical experiments we observe a real stability. However, we will discuss later on how much one can loose with respect to the solution of \eqref{backward-ter-inf-noL}.

For $n=N,\ldots,0$, the scheme \eqref{backward-ter-inf} gives back a function in the variables $(y,v,x)\in \mathcal{Y}\times\mathcal{V}_n\times\mathcal{X}_n$. Note that
$\mathcal{V}_n\times\mathcal{X}_n\subset I^V_n\times I^X_n$, where
$$
I^V_n=[v^n_0,v^n_n]\quad\mbox{and}\quad I^X_n=[x^n_0, x^n_n],
$$
that is, the intervals between the smallest and the biggest node at time-step $n$:
$$
v^n_0=\Big(\sqrt {V_0}-\frac{\sigma_V} 2\,n\sqrt{h}\Big)^2\I_{\{\sqrt {V_0}-\frac{\sigma_V} 2\,n\sqrt{h}>0\}},\quad
v^n_n=\Big(\sqrt {V_0}+\frac{\sigma_V} 2\,n\sqrt{h}\Big)^2,\quad
x^n_0=-n\sqrt{h},\quad
x^n_n=n\sqrt{h}.
$$
As $n$ decreases to 0, the intervals $I^V_n$ and $I^X_n$ are becoming smaller and smaller
and at time 0 they collapse to the single point $v^0_0=V_0$
and $x^0_0=X_0=0$ respectively. So, the norm we are going to define  takes into account these facts: at time $nh$ we consider for $\phi=\phi(t,y,v,x)$ the norm
\begin{equation}\label{norm}
\|\phi(nh,\cdot)\|_n = \sup_{(v,x)\in I^V_n\times I^X_n}\|\phi(nh,\cdot,v,x)\|_{l_2(\mathcal{Y})} = \sup_{(v,x)\in I^V_n\times I^X_n}\Big(\sum_{i\in\Z}|\phi(nh,y_i,v,x)|^2\Delta y\Big)^{\frac 12}.
\end{equation}
In particular,
\begin{align*}
&\|\phi(0,\cdot)\|_0 =\|\phi(0,\cdot,V_0,X_0)\|_{l_2(\mathcal{Y})} = \Big(\sum_{i\in\Z}|\phi(y_i,V_0,X_0)|^2\Delta y\Big)^{1/2}\quad\mbox{and}\\
&\|\phi(T,\cdot)\|_N
\leq \sup_{(v,x)\in\R_+\times \R} \|\phi(y_i,v,x)\|_{l_2(\mathcal{Y})}
= \sup_{(v,x)\in\R_+\times \R} \Big(\sum_{i\in\Z}|\phi(y_i,v,x)|^2\Delta y\Big)^{1/2}.
\end{align*}

We are now ready to give our stability result.

\begin{theorem}\label{prop-stability}
Let $f\geq 0$ and, in the case $g=f$, suppose that
$$
\sup_{t\in[0,T]}|f(t,y,v,x)|\leq \gamma_T|f(T,y,v,x)|,
$$
for some $\gamma_T>0$. Then, for every $L>0$ the numerical scheme \eqref{backward-ter-inf} is stable with respect to the norm \eqref{norm}:
$$
\|F^L_h(0,\cdot)\|_0\leq C^{N,L}_T
\|F^L_h(T,\cdot)\|_N
=C^{N,L}_T
\|f(T,\cdot)\|_N,\quad \forall h,\Delta y,
$$
where
$$
C^{N,L}_T=
\left\{
\begin{array}{ll}
e^{2\lambda cT+\sigma_rLT-\sum_{n=1}^N\varphi_{nh}h}\stackrel{N\to\infty}{\longrightarrow}
C^L_T=e^{2\lambda cT+\sigma_rLT-\int_0^T\varphi_{t}dt}&\mbox{ if } g=0,\smallskip\\
\max\Big(\gamma_T, e^{2\lambda cT+\sigma_rLT-\sum_{n=1}^N\varphi_{nh}h}\Big)\stackrel{N\to\infty}{\longrightarrow}
C^L_T=\max\Big(\gamma_T, e^{2\lambda cT+\sigma_rLT-\int_0^T\varphi_{t}dt}\Big)&\mbox{ if } g=f,
\end{array}
\right.
$$
in which $c>0$ is such that $\sum_l \nu(\xi_l)\Dy\leq \lambda c$.

\end{theorem}

\textbf{Proof.}
In order to weaken the notation, we set $g^n_{i,k,j}=g(nh,y_i,v^n_k,x^n_j)$ and, similarly, $F^n_{i,k,j}=F^L_h(nh,y_i,v^n_k,x^n_j)$, $(\mathfrak{I}_{a,b}F_h^{n+1})_{i,k_a,j_b}
=(\mathfrak{I}_{a,b}F^L_h)((n+1)h,y_i,v^{n+1}_{k_a (n,k)},x^{n+1}_{j_b (n,j)})$
(we have also dropped the dependence on $L$).
The scheme \eqref{backward-ter-inf} says that, at each time step $n<N$ and for each fixed $0\leq k,j\leq n$,
\begin{equation}\label{max}
F^n_{i,k,j} =
\max \Big\{g^n_{i,k,j},
e^{-(\sigma_r x^n_j1_{\{x^n_j>-L\}} + \varphi_{nh})h}u^{n}_{i,k,j} \Big\},
\end{equation}
where, according to \eqref{scheme_bates}, $u^{n}_{i,k,j}$ solves
\begin{equation}\label{scheme_bates1}
	\begin{array}{l} (\alpha_{n,k,j}-\beta_{n,k})u^n_{i-1,k,j}+(1+2\beta_{n,k})u^n_{i,k,j}-(\alpha_{n,k,j}+\beta_{n,k})u^n_{i+1,k,j}
	\smallskip\\
\displaystyle
	=\sum_{a,b\in\{d,u\}}p_{ab}(n,k,j) \Big((\mathfrak{I}_{a,b}F^{n+1})_{i,k_a,j_b}+h\Dy\sum_l\nu(\xi_l)
\big[(\mathfrak{I}_{a,b}F^{n+1})_{i+l,k_a,j_b}-(\mathfrak{I}_{a,b}F^{n+1})_{i,k_a,j_b}\big]\Big).
	\end{array}
\end{equation}
Let $\mathfrak{F}\varphi$ denote the Fourier transform of $\varphi\in l_2(\mathcal{Y})$, that is,
\begin{equation*}
  \mathfrak{F}\varphi(\theta) = \frac{\Dy}{\sqrt{2\pi}}\sum_{
	s\in\Z} \varphi_s e^{-\ii s\Delta y\theta},\quad \theta\in\R,
\end{equation*}
$\ii$ denoting the imaginary unit. We get from \eqref{scheme_bates1}
\begin{equation}\label{pp}
\begin{array}{l}
\Big((\alpha_{n,k,j}-\beta_{n,k})e^{-\ii \theta\Dy}+1+2\beta_{n,k}
-(\alpha_{n,k,j}+\beta_{n,k})e^{\ii \theta\Dy}\Big) \mathfrak{F} u^n_{k,j}(\theta)
\medskip\\
=\Big(1+h\Dy\sum_l\nu(\xi_l)(e^{\ii l\theta\Dy}-1)\Big)\sum_{a,b\in\{d,u\}}p_{ab}(n,k,j)
\mathfrak{F}(\mathfrak{I}_{a,b}F^{n+1})_{k_a,j_b}(\theta).
\end{array}
\end{equation}
Note that
\begin{align*}
|(\alpha_{n,k,j}-\beta_{n,k})&e^{-\ii \theta\Dy}+1+2\beta_{n,k}
-(\alpha_{n,k,j}+\beta_{n,k})e^{\ii \theta\Dy}|\\
&\geq\big|\mathfrak{Re}\big[(\alpha_{n,k,j}-\beta_{n,k})e^{-\ii \theta\Dy}
+1+2\beta_{n,k}-(\alpha_{n,k,j}+\beta_{n,k})e^{\ii \theta\Dy}\big]\big|\\
&=1+2\beta_{n,k}(1-\cos(\theta\Dy))
\geq 1,
\end{align*}
for every $\theta \in [0,2\pi)$ (recall that $\beta_{n,k}\geq 0$).
And since $\sum_l \nu(\xi_l)\Dy\leq \lambda c$, we obtain
\begin{align*}
| \mathfrak{F} u^n_{k,j}(\theta)|
&\leq \Big(1+h\Dy\sum_{l\in\Z} | e^{\ii l\theta\Dy}-1|\nu(\xi_l)\Big)
\sum_{a,b\in\{d,u\}}p_{ab}(n,k,j)|  \mathfrak{F}(\mathfrak{I}_{a,b}F^{n+1})_{k_a,j_b}(\theta)|\\
& \leq (1+2\lambda ch)\sum_{a,b\in\{d,u\}}p_{ab}(n,k,j)|  \mathfrak{F}(\mathfrak{I}_{a,b}F^{n+1})_{k_a,j_b}(\theta)|.
\end{align*}
Therefore,
\begin{align*}
\| \mathfrak{F} u^n_{k,j}\|_{L^2([0,2\pi),\mathrm{Leb})}
&\leq (1+2\lambda ch)\sum_{a,b\in\{d,u\}}p_{ab}(n,k,j)
\|\mathfrak{F}(\mathfrak{I}_{a,b}F^{n+1})_{k_a,j_b}\|_{L^2([0,2\pi),\mathrm{Leb})}.
\end{align*}
We use now the Parseval identity $\|\mathfrak{F} \varphi\|_{L^2([0,2\pi),\mathrm{Leb})}
=\|\varphi\|_{l_2(\mathcal{Y})}$ and we get
\begin{align*}
\| u^n_{\cdot, k,j}\|_{l^2(\mathcal{Y})}
&\leq (1+2\lambda ch)\sum_{a,b\in\{d,u\}}p_{ab}(n,k,j)
\|(\mathfrak{I}_{a,b}F^{n+1})_{\cdot, k_a,j_b}\|_{l^2(\mathcal{Y})}\\
&= (1+2\lambda ch)\sum_{a,b\in\{d,u\}}p_{ab}(n,k,j)
\|F^{n+1}_{\cdot, k_a,j_b}\|_{l^2(\mathcal{Y})},
\end{align*}
the first equality following from the fact that $i\mapsto (\mathfrak{I}_{a,b}F^{n+1})_{i, k_a,j_b}$ is a linear convex combination of translations of $i\mapsto F^{n+1}_{i, k_a,j_b}$ (see Remark \ref{interp}).
This gives
\begin{align*}
\sup_{0\leq k,j\leq n}\|e^{-(\sigma_r x^n_j\,1_{\{x^n_j>-L\}} + \varphi_{nh})h} u^n_{\cdot,k,j}\|_{l_2(\mathcal{Y})}
\leq (1+2\lambda c h)e^{\sigma_rLh-\varphi_{nh}h}
\sup_{0\leq k,j\leq n+1}\|F^{n+1}_{\cdot,k,j}\|_{l_2(\mathcal{Y})}
\end{align*}
and from \eqref{max}, we obtain
\begin{align*}
\sup_{0\leq k,j\leq n}\| F^n_{\cdot,k,j}\|_{l_2(\mathcal{Y})}
&\leq \max\Big(\sup_{0\leq k,j\leq n}\|g^n_{\cdot,k,j}\|_{l_2(\mathcal{Y})}, (1+2\lambda c h)e^{\sigma_rLh-\varphi_{nh}h}
\sup_{0\leq k,j\leq n+1}\|F^{n+1}_{\cdot,k,j}\|_{l_2(\mathcal{Y})}\Big).
\end{align*}
We now continue assuming that $g=f$, the case $g=0$ following in a similar way. So,
\begin{align*}
\sup_{0\leq k,j\leq n}\| F^n_{\cdot,k,j}\|_{l_2(\mathcal{Y})}
&\leq \max\Big(\gamma_T\|f(T,\cdot)\|_N, (1+2\lambda c h)e^{\sigma_rLh-\varphi_{nh}h}
\sup_{0\leq k,j\leq n+1}\|F^{n+1}_{\cdot,k,j}\|_{l_2(\mathcal{Y})}\Big).
\end{align*}
For $n=N-1$ we then obtain
\begin{align*}
\sup_{0\leq k,j\leq n}\| F^{N-1}_{\cdot,k,j}\|_{l_2(\mathcal{Y})}
&\leq \max\Big(\gamma_T\|f(T,\cdot)\|_N, (1+2\lambda c h)e^{\sigma_rLh-\varphi_{(N-1)h}h}
\|f(T,\cdot)\|_{N}\Big)
\end{align*}
and by iterating the above inequalities, we finally get
\begin{align*}
\| F^0\|_0
=\|F^0_{\cdot,0,0}\|_{l_2(\mathcal{Y})}
&\leq \max\Big(\gamma_T\|f(T,\cdot)\|_N, (1+2\lambda c h)^Ne^{N\sigma_rLh-\sum_{n=1}^N\varphi_{nh}h}
\|f(T,\cdot)\|_N\Big).
\end{align*}
\cvd

\begin{remark}\label{antitrasf}
In the above proof we have actually proved that, as $n$ varies, the solution $u^n_{\cdot,k,j}$ to the infinite linear system \eqref{scheme_bates} does exist and is unique if $\|f(T,\cdot)\|_N<\infty$. In fact, starting from equality \eqref{pp}, we define the function $\psi_{k,j}(\theta)$, $\theta\in[0,2\pi)$, by
$$
\begin{array}{l}
\Big((\alpha_{n,k,j}-\beta_{n,k})e^{-\ii \theta\Dy}+1+2\beta_{n,k}
-(\alpha_{n,k,j}+\beta_{n,k})e^{\ii \theta\Dy}\Big) \psi_{k,j}(\theta)
\smallskip\\
=\Big(1+h\Dy\sum_l\nu(\xi_l)(e^{\ii l\theta\Dy}-1)\Big)\sum_{a,b\in\{d,u\}}p_{ab}(n,k,j)
\mathfrak{F}(\mathfrak{I}_{a,b}F^{n+1})_{k_a,j_b}(\theta).
\end{array}
$$
As noticed in the proof of Proposition \ref{prop-stability}, the factor multiplying $\psi_{k,j}(\theta)$ is different from zero because $\beta_{n,k}\geq 0$. So, the definition of $\psi_{k,j}$ is well posed and moreover, $\psi_{k,j}\in L^2([0,2\pi,),\mathrm{Leb})$. We now set $u^n_{\cdot,k,j}$ as the inverse Fourier transform of $\psi_{k,j}$, that is,
$$
u^n_{l,k,j}=\frac 1{\Delta y\sqrt{2\pi}}\int_0^{2\pi}\psi_{k,j}(\theta)e^{\mathbf{i}\,l\theta\Delta y}d\theta,\quad l\in\Z.
$$
Straightforward computations give that $u^n_{\cdot,k,j}$ fulfils the equation system \eqref{scheme_bates}.
\end{remark}

Of course, Theorem \ref{prop-stability} gives a stability property for the scheme introduced in \cite{bcz-hhw} for the Heston-Hull-White model (just take $\lambda=0$ - no jumps are considered). Moreover, for the standard Bates model, that is, $\sigma_r=0$ (deterministic interest rate),  Theorem \ref{prop-stability} applies to the original (untruncated) scheme  \eqref{backward-ter-inf-noL}.

\subsection{Back to the original scheme \eqref{backward-ter-inf-noL}}

Let us now discuss what may happen when one introduces the threshold $L$. We recall that the original scheme \eqref{backward-ter-inf-noL} gives the numerical approximation of the function $F$ in \eqref{F} whereas
the discount truncated scheme \eqref{backward-ter-inf} aims to numerically compute the function $F^L$ in \eqref{FL}. Proposition \ref{prop-FL} below shows that, under standard hypotheses, $F^L$ tends to $F$ as $L\to\infty$ very fast. This means that, in practice, we loose very few in using \eqref{backward-ter-inf} in place of \eqref{backward-ter-inf-noL}.
\begin{proposition}\label{prop-FL}
Suppose that $f=f(t,y,v,x)$ has a polynomial growth in the variables $(y,v,x)$, uniformly in $t\in[0,T]$. Let $F$ and $F^L$, with $L>0$, be defined in \eqref{F} and \eqref{FL} respectively. Then there exist positive constants $c_T$ and $C_T(y,v,x)$ 
 such that for every $L>0$
$$
|F(t,y,v,x)-F^L(t,y,v,x)|
\leq \sigma_r C_T(y,v,x)
e^{-c_T|L+xe^{-\kappa_r(T-t)}|^2},
$$
for every $t\in[0,T]$ and $(y,v,x)\in\R\times\R_+\times \R$.
\end{proposition}

\textbf{Proof. }
In the following, $C$ denotes a positive constant, possibly changing from line to line. We have
\begin{align*}
&|F(t,y,v,x)-F^L(t,y,v,x)|\\
&\leq \sigma_r C\E\Big(\sup_{t\leq u\leq T}|f(u,Y_u^{t,y,v,x}, V_u^{t,v}, X^{t,v})|
\times e^{-\sigma_r\int_t^uX_s^{t,x}1_{\{X_s^{t,x}>-L\}}ds}\times
\Big|e^{-\sigma_r\int_t^uX_s^{t,x}1_{\{X_s^{t,x}<-L\}}ds}-1\Big|
\Big).
\end{align*}
Set 
$$
\tau_{-L}^{t,x}=\inf\{s\geq t\,:\, X_s^{t,x}\leq -L\}.
$$
One has $1_{\{X_s^{t,x}<-L\}}\leq 1_{\{\tau_{-L}^{t,x}< s\}}$ so that, for $u\leq T$,
$$
0\leq -\sigma_r\int_t^uX_s^{t,x}\,1_{\{X_s^{t,x}<-L\}}ds
\leq \sigma_r\int_t^u|X_s^{t,x}|\,1_{\{\tau_{-L}^{t,x}< s\}}ds
\leq \sigma_r\int_{t}^T|X_s^{t,x}|ds\,1_{\{\tau_{-L}^{t,x}\leq T\}}.
$$
Then,
\begin{align*}
&\sup_{t\leq u\leq T}e^{-\sigma_r\int_t^uX_s^{t,x}1_{\{X_s^{t,x}>-L\}}ds}
\Big|e^{-\sigma_r\int_t^uX_s^{t,x}1_{\{X_s^{t,x}<-L\}}ds}-1\Big|\\
&\qquad\leq 
e^{\sigma_r\int_{t}^T|X_s^{t,x}|ds}\Big(e^{\sigma_r\int_{t}^T|X_s^{t,x}|ds}-1\Big)\,1_{\{\tau_{-L}^{t,x}\leq T\}}
\leq 
2e^{2\sigma_r\int_{t}^T|X_s^{t,x}|ds}\,1_{\{\tau_{-L}^{t,x}\leq T\}}
\end{align*}
By inserting,
\begin{align*}
&|F(t,y,v,x)-F^L(t,y,v,x)|\\
&\leq \sigma_rC\E\Big(\sup_{t\leq u\leq T}|f(u,Y_u^{t,y,v,x}, V_u^{t,v}, X_u^{t,v})|
\,e^{2\sigma_r\int_t^T|X_s^{t,x}|ds}\,1_{\{\tau_{-L}^{t,x}\leq T\}}\Big)\\
&\leq \sigma_rC\E\Big(\sup_{t\leq u\leq T}|f(u,Y_u^{t,y,v,x}, V_u^{t,v}, X_u^{t,v})|^2
\,e^{4\sigma_r\int_t^T|X_s^{t,x}|ds}\Big)^{1/2}
\P(\tau_{-L}^{t,x}\leq T)^{1/2}.
\end{align*}
Since $f$ has a polynomial growth in the space variables, uniformly in the time variable, by standard estimates one gets that 
$\sup_{t\leq u\leq T}|f(u,Y_u^{t,y,v,x}, V_u^{t,v}, X^{t,v})|$ has all moments. Moreover, for a Brownian motion $W$, $\sup_{0<s<T}|W_s|$ has finite exponential moments of any order, and this gives that  $e^{4\sigma_r\int_t^T|X_s^{t,x}|ds}$ has finite moments of any order. It follows that
\begin{align*}
&|F(t,y,v,x)-F^L(t,y,v,x)|
\leq C
\P(\tau_{-L}^{t,x}\leq T)^{1/2}.
\end{align*}
As for the above probability, recall that $X_s^{t,x}=xe^{-\kappa_r(s-t)}+\int_t^se^{-\kappa_r(s-u)}dW^2_u$ so that
\begin{align*}
&\P(\tau_{-L}^{t,x}\leq T)
=\P(\inf_{s\in[t,T]}X_s^{t,x}<-L)
=\P\Big(\inf_{s\in[t,T]}\Big(xe^{-\kappa_r(s-t)}+\int_t^se^{-\kappa_r(s-u)}dW^2_u\Big)<-L\Big)\\
&\leq \P\Big(\sup_{s\in[t,T]}\Big|\int_t^se^{\kappa_ru}dW^2_u\Big|>L+xe^{-\kappa_r(T-t)}\Big)
\leq 2\exp\Big(-\frac{|L+xe^{-\kappa_r(T-t)}|^2}{2\int_t^Te^{2\kappa_ru}du} \Big).
\end{align*}
By inserting above, we get the result.
\cvd

\subsection{Further remarks}

As already stressed, the introduction of the threshold $-L$ allows one to handle the discount term. In order to let the discount disappear, an approach consists in the use of a transformed function, as  developed by several authors (see e.g. Haentjens and in't Hout \cite{Hint} and references therein). This is a nice fact for European options (PIDE problem), being on the contrary a non definitive tool when dealing with American options (obstacle PIDE problem). Let us see why.

First of all, let us come back to the model for the triple $(Y,V,X)$, see \eqref{YVX-dyn}. The infinitesimal generator is
\begin{equation}\label{Lt}
\begin{array}{ll}
L_tu
=&
\displaystyle
\Big(\sigma_rx+\varphi_t-\eta-\frac 12v\Big)\partial_yu
+\kappa_V(\theta_V-v)\partial_vu
-\kappa_rx\partial_xu\smallskip\\
&\displaystyle
+\frac 12\Big(v\partial^2_{yy}u+\sigma_V^2v\partial^2_{vv}u+\partial^2_{xx}u
+2\rho_1\sigma_Vv\partial^2_{yv}u+2\rho_2\sqrt  v\,\partial^2_{yx}u\Big)\smallskip\\
&\displaystyle
+ \displaystyle\int_{-\infty}^{+\infty} \left[ u(t,y+\xi;v,x)-u(t,y;v,x)\right]\nu(\xi) d\xi.
\end{array}
\end{equation}
We set
$$
G(t,x)=\E\Big(e^{-\sigma_r\int_t^T X^{t,x}_s ds}\Big).
$$
Recall that (see e.g. \cite{ll})
\begin{equation}\label{G}
G(t,x)=e^{-x\sigma_r\Lambda(t,T)-\frac {\sigma_r^2}{2\kappa_r^2}(\Lambda(t,T)-T+t)-\frac {\sigma_r^2}{4\kappa_r}\Lambda^2(t,T)},\quad \Lambda(t,T)=\frac{1-e^{-\kappa_r (T-t)}}{\kappa_r}
\end{equation}
and, moreover, $G$ solves the PDE
\begin{equation}\label{PDE-G}
\begin{array}{l}
\displaystyle
\partial_tG
-\kappa_r x\partial_x G+\frac 12\partial_{xx}^2 G-\sigma_rxG=0,\quad t\in[0,T), x\in\R,\smallskip\\
G(T,x)=1.
\end{array}
\end{equation}

\begin{lemma}\label{G-lemma}
Let $L_t$ denote the infinitesimal generator in \eqref{Lt}. Set $\overline{u}=u\cdot G^{-1}$. Then
$$
\partial_tu +L_tu -xu= G\big(\partial_t\bu +\bL_t \bu \big),
$$
where
$$
\bL_t=L_t-\sigma_r\frac{1-e^{-\kappa_r (T-t)}}{\kappa_r}\big[\rho_{2}\sqrt v \partial_{y}+\partial_x\big].
$$
\end{lemma}

\textbf{Proof.}
Since $G$ depends on $t$ and $x$ only, straightforward computations give
\begin{align*}
\partial_tu +L_tu -xu=&
G\big[\partial_t \bu +L_t\bu\big]
+\partial_xG(t,x)
\big[\rho_{2}\sqrt v \partial_{y}\bu+\partial_x\bu\big]
+\bu\big[\partial_tG-\kappa_rx\partial_x G+\frac 12\partial_{xx}^2 G-\sigma_rxG\big].
\end{align*}
By \eqref{PDE-G}, the last term is null.  The statement now follows by observing that $\partial_x\ln G(t,x)=-\sigma_r\frac{1-e^{-\kappa_r (T-t)}}{\kappa_r}$. \cvd

\medskip

We notice that the operator $\overline{L}_t$ in Lemma \ref{G-lemma} is the infinitesimal generator of the jump-diffusion process $(\overline{Y},\overline{V},\overline{X})$ which solves the stochastic differential equation as in \eqref{YVX-dyn},  with the same diffusion coefficients and jump-terms but with the new drift coefficients
$$
\mu_{\overline{Y}}(t,v,x)=\mu_Y(v,x)-\sigma_r\frac{1-e^{-\kappa_r (T-t)}}{\kappa_r}
\rho_{2}\sqrt v,\quad
\mu_{\overline{V}}(v)\equiv \mu_V(v),
\quad
\mu_{\overline{X}}(x)=\mu_X(t,x)-\sigma_r\frac{1-e^{-\kappa_r (T-t)}}{\kappa_r}.
$$
Let us first discuss the scheme \eqref{backward-ter-inf-noL} for $F$ in the case $g=0$ (European options). By passing to the associated PIDE,  Lemma \ref{G-lemma} says that
$$
F(t,y,v,x)
=G(t,x)\overline{F}(t,y,v,x),
$$
where
$$
\overline{F}(t,y,v,x)
=\E(e^{-\int_t^T\varphi_sds}f(T,\overline{Y}_T^{t,y,v,x},\overline{V}_T^{t,v},\overline{X}_T^{t,x})).
$$
Therefore, in practice one has to numerically evaluate the function $\overline{F}$. By using our hybrid tree/finite-difference approach, this means to consider the scheme in \eqref{backward-ter-inf}, with the new coefficient $\overline{\alpha}_{n,k,j}$ (written starting from the new drift coefficients)  but with a discount depending on the (deterministic) function $\varphi$ only, that is, with $e^{-(\sigma_rx^n_j1_{\{x^n_j>-L\}}+\varphi_{nh})h}$ replaced by $e^{-\varphi_{nh}h}$. And the proof of the Proposition \ref{prop-stability} shows that one gets
$$
\|\overline{F}_h(0,\cdot)\|_0\leq \max\big(\gamma_T,e^{2\lambda cT-\sum_{n=0}^N\varphi_{nh}h}\big)\|f(T,\cdot)\|_N.
$$
In other words, by using a suitable transformation, the European scheme is always stable and  no thresholds are needed.

Let us discuss now the American case, that is, the scheme  \eqref{backward-ter-inf-noL} with $g=f$. One could  think to use the above transformation in order to get rid of the exponential depending on the process $X$. Set again
$$
\overline{F}(t,y,v,x)
=G(t,x)^{-1}F(t,y,v,x).
$$
By using the associated obstacle PIDE problem, Lemma \ref{G-lemma} suggests that
$$
\overline{F}(t,y,v,x)
=\sup_{\tau\in\mathcal{T}_{t,T}}\E(e^{-\int_t^\tau \varphi_sds}\overline{f}(\tau,\overline{Y}_\tau^{t,y,v,x},\overline{V}_\tau^{t,v},\overline{X}_\tau^{t,x})),
\quad \mbox{with}\quad
\overline{f}(t,y,v,x)=G^{-1}(t,x)f(t,y,v,x).
$$
So, in order to numerically compute $\overline{F}$, one needs to set up the scheme \eqref{backward-ter-inf} with the new coefficient $\overline{\alpha}_{n,k,j}$, with $f$ replaced by $\overline f$, $g=\overline{f}$ and with the discounting factor  $e^{-(\sigma_rx^n_j1_{\{x^n_j>-L\}}+\varphi_{nh})h}$ replaced by $e^{-\varphi_{nh}h}$. So, again one is able to cancel the unbounded part of the discount. Nevertheless, the unpleasant point is that even if $\|f(T,\cdot)\|_N$ has a bound which is uniform in $N$ then $\|\overline{f}(T,\cdot)\|_N$ may have not because $G^{-1}(t,x)$ has an exponential containing $x$, see \eqref{G}. In other words, the unboundedness problem appears now in the obstacle.

\section{The hybrid Monte Carlo and tree/finite-difference approach algorithms in practice}\label{practice}

The present section is devoted to our numerical experiments. We first resume the main steps of our algorithms and then we present several numerical tests.

\subsection{A schematic sketch of the main computational steps in our algorithms}\label{sec:pseudoalg}
To summarize, we resume here the main computational steps of the two proposed algorithms.

First, the procedures need the following preprocessing steps, concerning the construction of the bivariate tree:
\begin{itemize}
	\item[(\textsc{T1})] define a discretization of $[0,T]$ in $N$ subintervals $[nh,(n+1)h]$, $n=0,\ldots,N-1$, with $h=T/N$;
	\item[(\textsc{T2})] for the process $V$, set the binomial tree $v^n_k$, $0\leq k \leq n\leq N$, by using
	\eqref{state-space-V}, then compute the jump nodes $k_a (n,k)$
	and the jump probabilities $p^{V}_a(n,k)$, $a\in\{u,d\}$,  by using  \eqref{ku}-\eqref{kd} and \eqref{pik};
	\item[(\textsc{T3})] for the process $X$, set the binomial tree $x^n_j$, $0\leq j\leq N$, by using \eqref{state-space-X}, then compute the jump nodes $j_b (n,j)$ and the jump probabilities $p^{X}_b(n,j)$, $b\in\{u,d\}$,   by using \eqref{ju}-\eqref{jd} and \eqref{pij};
	\item[(\textsc{T4})] for the $2$-dimensional process $(V,X)$, merge the binomial trees in the bivariate tree $(v^n_k,x^n_j)$, $0\leq k,j\leq n\leq N$, by using \eqref{state-space-Vr}, then compute the jump-nodes $(k_a (n,k),j_b (n,j))$ and the transitions probabilities $p_{ab}(n,k,j)$, $(a,b)\in\{d,u\}$, by using \eqref{treescheme}.
\end{itemize}
The bivariate tree for $(V,X)$ is now settled. Our hybrid tree/finite-difference algorithm can be resumed as follows:

\begin{enumerate}
	\item[(\textsc{FD1})] set a mesh grid $y_i$ for the solution of all the PIDEs;
	\item[(\textsc{FD2})] for each node $(v_{N,k},
	x_{N,j})$, $0\leq k,j\leq N$, compute the option prices at maturity for each $y_i$, $i\in\mathcal{Y}_M$, by using the payoff function;
	\item[(\textsc{FD3})] for $n=N-1,\ldots 0$: for each $(v^n_k,x^n_j)$, $0\leq k,j\leq n$, compute the option prices for each $y_i\in \mathcal{Y}_M$, by solving the linear system \eqref{u-interp}.
\end{enumerate}

Notice that, at each time step $n$,  we need only the one-step PIDE solution in the time interval $[nh, (n+1)h]$. Moreover, both the (constant) PIDE coefficients and the Cauchy final condition change according to the position of the volatility and the interest rate components on the bivariate tree at time step $n$.

We conclude by briefly recalling the main steps of the hybrid Monte Carlo method:
\begin{enumerate}
	\item[(\textsc{MC1})]
	let the chain $(\hat V^h_n,\hat X^h_n)$ evolve for $n=1,\ldots,N$, following the probability structure in (\textsc{T4});
	\item[(\textsc{MC2})]
	generate $\Delta_1,\ldots,\Delta_N$ i.i.d. standard normal r.v.'s independent of the noise driving the chain $(\hat V^h,\hat X^h)$;
	\item[(\textsc{MC3})] generate $K_h^1,\ldots,K_h^N$ i.i.d. positive Poisson r.v.'s of parameter $\lambda h$,  independent of both the chain $(\hat V^h,\hat X^h)$ and the Gaussian r.v.'s $\Delta_1,\ldots,\Delta_N$, and for every $n=1,\ldots,N$, if $K_h^n>0$ simulate the corresponding amplitudes $\log(1+J_1^n),\ldots ,\log(1+J^n_{K_h^n})$;
	\item[(\textsc{MC4})] starting from $\hat Y_0^h=Y_0$, compute the approximate values $\hat Y^h_n$, $1\leq n \leq N$, by using \eqref{hmc};
	\item[(\textsc{MC5})] following the desired Monte Carlo
	method (European or Longstaff-Schwartz algorithm \cite{ls} in the case of American options), repeat the above simulation scheme and compute the option price.
\end{enumerate}

\begin{remark}\label{standard-bates}
	In next Section \ref{sect-numerics} we develop numerical experiments also in the standard Bates model, that is, with a constant interest rate.
Recall that in the standard Bates model the dynamic reduces to
	\begin{equation}\label{stand-bates}
	\begin{array}{l}
	\displaystyle\frac{dS_t}{S_{t^-}}= (r-\eta)dt+\sqrt{V_t}\, dZ^S_t+d H_t,
	\smallskip\\
	dV_t= \kappa_V(\theta_V-V_t)dt+\sigma_V\sqrt{V_t}\,dZ^V_t,
	\end{array}
	\end{equation}
	with $S_0,V_0>0$, $r\geq 0$ constant parameters, $d\<Z^S,Z^V\>_t=\rho dt,\quad |\rho|<1$ and
	$H_t$ is the compound Poisson process already introduced in Section \ref{sect-model}, see \eqref{H}. We can apply our hybrid approach to this case as well: it suffices just to follow the computational steps listed above except for the construction of the binomial tree for the process $X$. Consequently, we do not need the bivariate tree for $(V,X)$, specifically we omit steps \textsc{(T3)}-\textsc{(T4)} and we replace step \textsc{(MC1)} with
	\begin{enumerate}
		\item[\textsc{(MC1')}]
		let the chain $\hat V^h_n$ evolve for $n=1,\ldots,N$, following the probability structure in \textsc{(T2)}.
	\end{enumerate}
	And of course, in all computations we set equal to 0 the parameters involved in the dynamics for $r$, except for the starting value $r_0$. In particular, we have $\sigma_r=0$ and $\varphi_t=r_0$ for every $t$.
\end{remark}

\begin{remark} \label{costo}
	We observe that in order to compute the option price by the hybrid tree/finite-difference procedure, in step \textsc{(FD3)} we need to solve many times the tridiagonal system \eqref{u-interp}. This is typically solved by the LU-decomposition method in $O(M)$ operations (recall that the total number of the grid values $y_i\in\mathcal{Y}_M$ is $2M+1$).
	However, due to the approximation of the integral term \eqref{num_int}, at each time step $n<N$ we have to compute the sum
	\begin{equation}\label{int_sum}
	\sum \tilde u^{n+1}_{i+l} \nu(\xi_l),
	\end{equation}
	which is the most computationally expensive step of this part of the algorithm: when applied directly, it requires $O(M^2)$ operations. Following the Premia software implementation \cite{pr}, in our numerical tests we use the Fast Fourier Transform  to compute the term \eqref{int_sum} and the computational costs of this step reduce to $O(M\log M)$.  According to the Bates (respectively the Bates-Hull-White) model, the hybrid algorithm requires $N(N+1)/2$  (respectively $(N(N+1)/2)^2$) resolutions of linear systems, each of them having a  linear complexity. Therefore, the overall complexity is $O(N^2M\log M)$ (respectively $O(N^4M\log M)$).
	
\end{remark}

\subsection{Numerical results}\label{sect-numerics}

We develop several numerical results in order to asses the
efficiency and the robustness of the hybrid tree/finite-difference
method and the hybrid Monte Carlo method in the case of
plain vanilla options. The Monte Carlo results derive from our hybrid simulations and, for American options, the use of the Monte Carlo algorithm by Longstaff and Schwartz in \cite{ls}.

We first provide results for the standard Bates model (see Remark \ref{standard-bates}) and secondly, for the case in which the interest rate process is assumed to be stochastic, see \eqref{BHHmodel}.

Following Chiarella \emph{et al.} \cite{ckmz}, in our numerical tests we assume that the jumps for the log-returns are normal, that is,
\begin{equation}\label{nuC}
\log(1+J_1)\sim N\Big(\gamma-\frac 12\delta^2,\delta^2\Big),
\end{equation}
$N$ denoting the Gaussian law (we also notice that the results in \cite{ckmz} correspond to the choice $\gamma=0$).
In Section \ref{sect-num-bates}, we first compare our results with the ones provided in Chiarella \emph{et al.}
\cite{ckmz}. Then in Section \ref{andersen} we study options with large maturities and when the Feller condition is not fulfilled. Finally, Section \ref{sect-num-bateshw} is devoted to  test experiments for European and American options in the Bates model with stochastic interest rate. The codes have been written by using the C++ language and the computations have been all performed in double precision on a PC 2,9 GHz Intel Core I5 with 8 Gb of RAM.

\subsubsection{The standard Bates model}\label{sect-num-bates}

We refer here to the standard Bates model as in \eqref{stand-bates}.
In the European and American option contracts we are dealing with, we
consider the following set of parameters, already used in the numerical results
provided in Chiarella \emph{et al.} \cite{ckmz}: 
initial price $S_0=80,
	90, 100, 110, 120$, strike price $K=100$,
	maturity $T=0.5$; (constant) interest rate $r=0.03$, dividend rate $\eta=0.05$; initial
	volatility $V_0=0.04$, long-mean $\theta_V=0.04$, speed of
	mean-reversion   $\kappa_V=2$, vol-vol $\sigma_V=0.4$,
	correlation $\rho=-0.5,0.5$; intensity $\lambda=5$, jump parameters $\gamma=0$ and $\delta=0.1$ (recall \eqref{nuC}).
It is known that the case $\rho>0$ may lead to moment explosion, see. e.g. \cite{ap}. Nevertheless, we report here results for this case as well, for the sake of  comparisons with the study in Chiarella \textit{et al.} \cite{ckmz}.

In order to numerically solve the PIDE  using the finite difference
scheme, we first localize the variables and the integral term to
bounded domains. We use for this purpose the estimates for the
localization domain and the truncation of large jumps given by Voltchkova and Tankov
\cite{vota08}. For example, for the previous model parameters the PIDE
problem is solved in the finite interval
$[\ln S_0-1.59, \ln S_0+1.93]$.

The numerical study of the hybrid tree/finite-difference method
\textbf{HTFD} is split in two cases
\begin{itemize}
	\item[]
	\textbf{HTFDa}: time steps $N_t=50$ and varying mesh grid $\Delta
	y=0.01$, $0.005$, $0.0025$, $0.00125$;
	\item[]
	\textbf{HTFDb}: time steps $N_t=100$ and varying mesh grid $\Delta
	y=0.01$, $0.005$, $0.0025$, $0.00125$.
\end{itemize}

Concerning the Monte Carlo method, we give  the results from the hybrid simulation scheme in Section \ref{sect-MC}, that we call \textbf{HMC}. We give comparisons with the accurate third-order Alfonsi \cite{al} discretization scheme for the CIR stochastic volatility process and by using an exact scheme for the
interest rate. In addition, we simulate the jump component in the standard way. The resulting Monte Carlo scheme is here called \textbf{AMC}.
We consider  varying number of Monte Carlo
iterations $N_{\mathrm{MC}}$ and two cases for the number of time discretization steps
iterations:
\begin{itemize}
	\item[]
	\textbf{HMCa} and \textbf{AMCa}: $N_t=50$ and $N_{\mathrm{MC}}=10000, 50000, 100000, 200000$;
	\item[]
	\textbf{HMCb} and \textbf{AMCb}: $N_t=100$ and $N_{\mathrm{MC}}=10000, 50000, 100000, 200000$.
\end{itemize}
All Monte Carlo results include the associated $95\% $ confidence
interval.

Table \ref{tab1} reports European call option
prices. Comparisons are
given with a  benchmark value obtained using the Carr-Madan pricing
formula \textbf{CF} in \cite{CM} that applies Fast Fourier Transform
methods (see the Premia software implementation \cite{pr}).

In Table \ref{tab2} we provide results for
American call option prices. In this case we compare with the values obtained by using
the method of lines in \cite{ckm}, called \textbf{MOL}, with mesh
parameters $200$ time-steps,
$250$ volatility lines, $2995$ asset grid points, and the \textbf{PSOR} method with mesh parameters $1000, 3000, 6000$
that Chiarella \emph{et al.} \cite{ckmz} used as the true solution.
Moreover, we consider the Longstaff-Schwartz \cite{ls} Monte Carlo
algorithm both for \textbf{AMC} and \textbf{HMC}. In particular
\begin{itemize}
	\item[]
	\textbf{HMCLSa} and \textbf{AMCLSa}: $10$ exercise dates, $N_t=50$ and $N_{\mathrm{MC}}=10000, 50000, 100000, 200000$;
	\item[]
	\textbf{HMCLSb} and \textbf{AMCLSb}: $20$ exercise dates, $N_t=100$ and $N_{\mathrm{MC}}=10000, 50000, 100000, 200000$.
\end{itemize}

Tables \ref{tab31} and \ref{tab32} refer to the computational time cost (in seconds) of
the various algorithms for $\rho=-0.5$ in the European and
American case respectively.

In order to study the convergence behavior of  our approach \textbf{HTFD},  we consider the convergence ratio proposed in \cite{dfl}, defined as
\begin{equation}\label{ratio}
\mathrm{ratio}=\frac{P_{\frac{N}{2}}-P_{\frac{N}{4}}}{P_{N}-P_{\frac{N}{2}}},
\end{equation}
where $P_{N}$ denotes here the approximated price obtained with
$N=N_t$ number of  time steps. Recall that $P_{N}=O(N^{-\alpha})$ means that
$\mathrm{ratio}=2^{\alpha}$.
Table \ref{tab4ratio} suggests that the convergence ratio for
\textbf{HTDFb} is approximatively linear.

The numerical results in Table \ref{tab1}-\ref{tab32}
show that \textbf{HTFD} is accurate, reliable and efficient for pricing European and American
options in the Bates model.
Moreover, our hybrid Monte Carlo algorithm \textbf{HMC} appears to be competitive with \textbf{AMC}, that is the one from the  simulations by Alfonsi \cite{al}:
the numerical results are similar in term of precision
and variance but \textbf{HMC} is definitely  better from the computational times point of view. Additionally, because of its simplicity,
\textbf{HMC} represents a real and interesting alternative to
\textbf{AMC}.

\begin{figure}[ht]
	\begin{center}
		\includegraphics[scale=0.35]{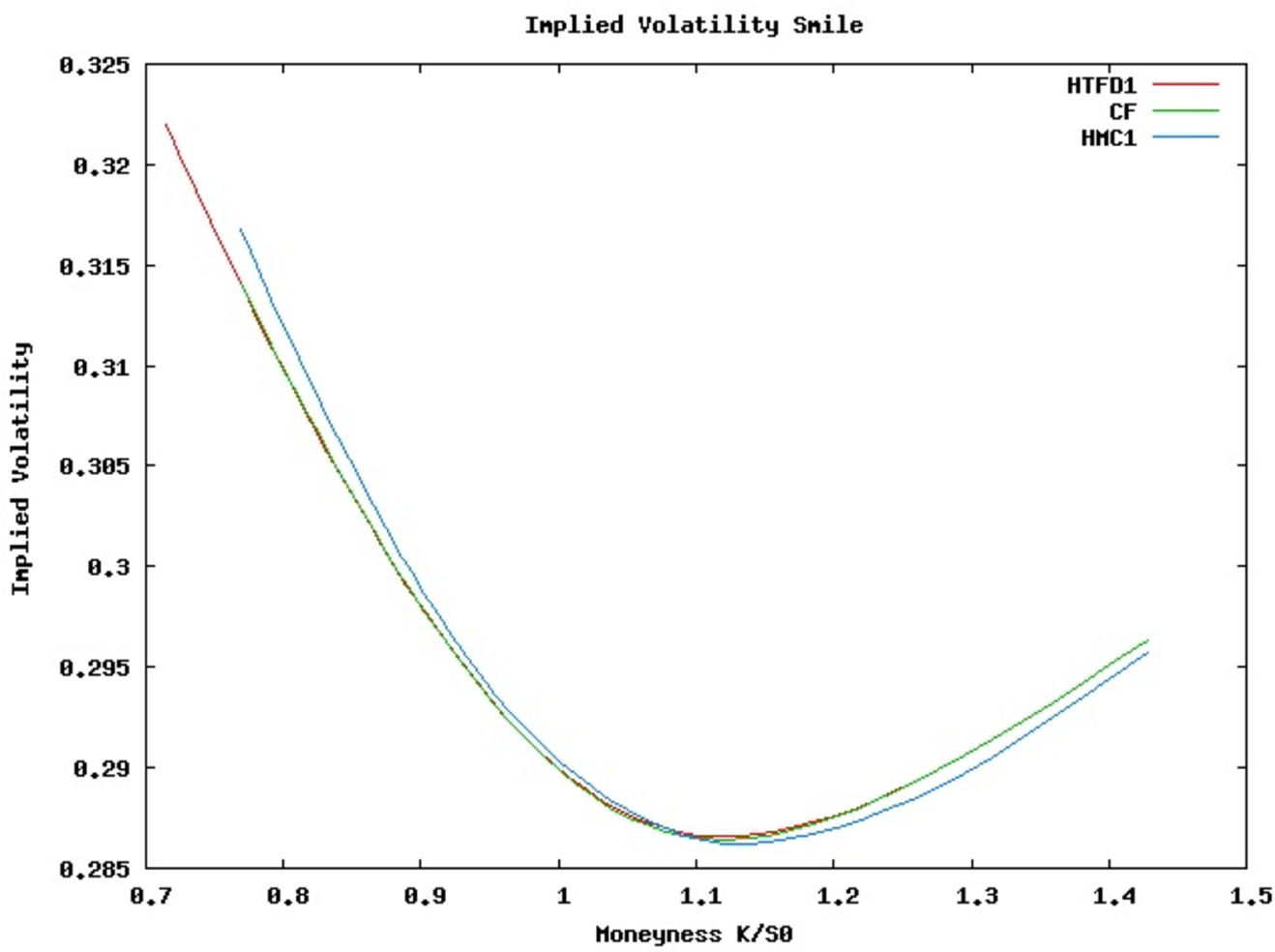}
		\caption{\em \small Standard Bates model. Moneyness vs implied volatility for European call options with $N_t=50$. The red line refers to \textbf{HTFDa} with
			$\Delta y=0.005$, the blue line refers to \textbf{HMCa} with
			$N_{\mathrm{MC}}=50000$ and the green line refers to the benchmark values \textbf{CF}. The accuracy of our hybrid methods is evident.
		}    
		\label{Fig1}
	\end{center}
\end{figure}
\begin{figure}[ht]
	\begin{center}
		\includegraphics[scale=0.35]{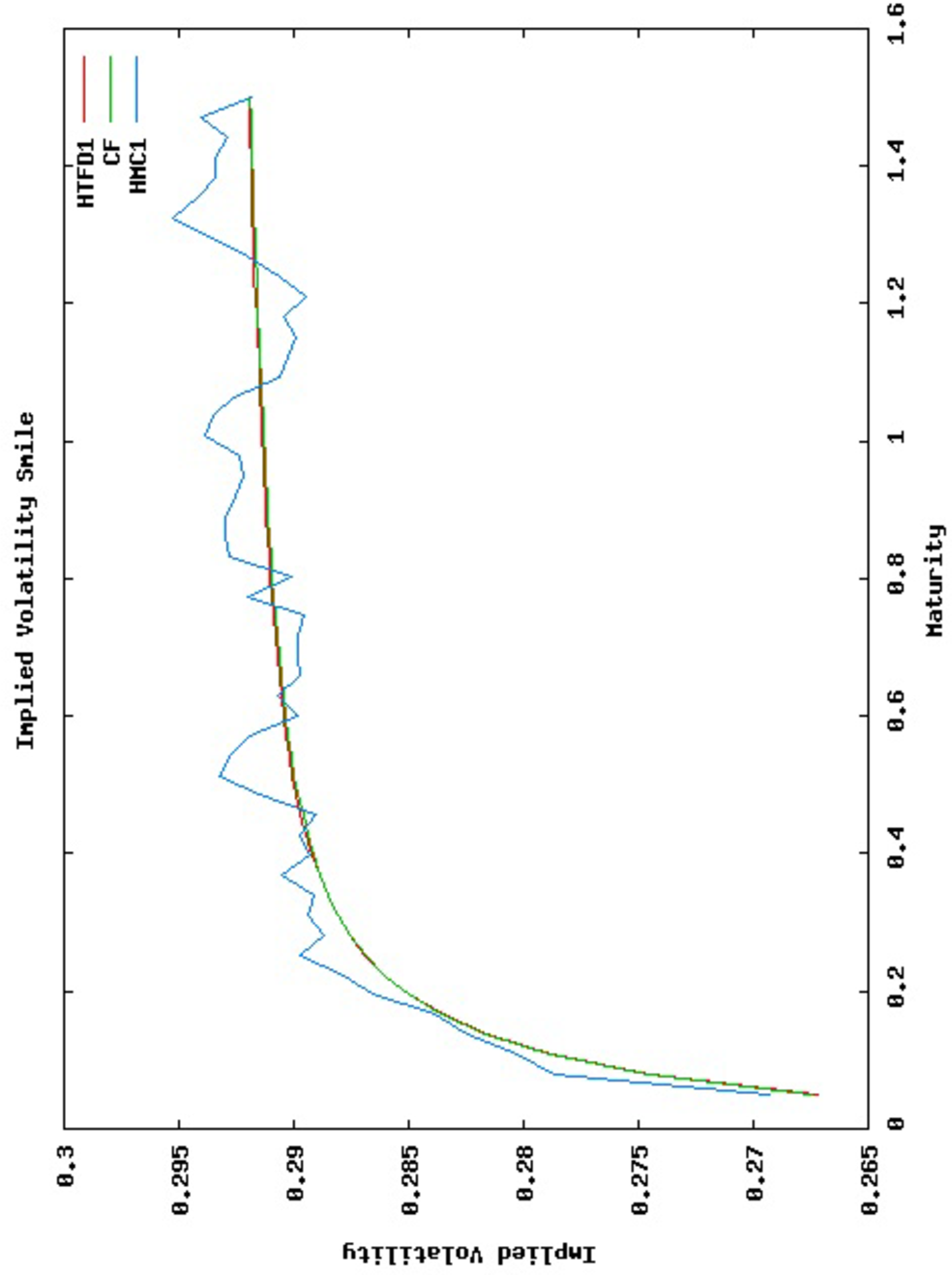}
		\caption{\em\small Standard Bates model. Maturity vs implied volatility for European call options with $N_t=50$. 
			The red line refers to \textbf{HTFDa} with
			$\Delta y=0.005$, the blue line refers to \textbf{HMCa} with
			$N_{\mathrm{MC}}=50000$, and the green line refers to the benchmark values \textbf{CF}. Also in this case the accuracy of our hybrid methods is evident. 
			%
			%
			%
		}
		\label{Fig2}
	\end{center}
\end{figure}

As a further evidence of the accuracy of our hybrid methods,
	in Figure \ref{Fig1} and \ref{Fig2} we study the shapes of implied volatility smiles across
	moneyness $\frac{K}{S_0}$ and maturities $T$ using \textbf{HTFDa} and \textbf{HMCa}. We compare the
	graphs with the results from the benchmark values \textbf{CF}. The parameters used for these two tests cases are:
	$T=0.5$, $S_0=100$, $K=100$,
	$r=0.03$, $\eta=0.05$, $V_0=0.04$, $\theta_V=0.04$, $\kappa_V=2$, $\sigma_V=0.4$,
	$\lambda=5$, $\gamma=0$, $\delta=0.1$, $\rho=-0.5$.

\begin{table}[ht]
	\centering
	\subtable[]{
		\footnotesize
		{\begin{tabular} {@{}c|lrr|c|rrrrrc@{}} 
				\toprule $\rho=-0.5$ & $\Delta y$
				&HTFDa  &HTFDb & CF &$N_{\mathrm{MC}} $&HMCa  &HMCb &AMCa  &AMCb\\
				\hline
				& 0.01& 1.1302& 1.1302 & &10000&1.08$\pm$0.09 &1.11$\pm$0.09
				&1.00$\pm$0.09&1.08$\pm$0.09 &\\
				&0.005& 1.1293& 1.1294 & &50000 &1.12$\pm$0.04 &1.17$\pm$0.04 &1.07$\pm$0.04&1.10$\pm$0.04 &\\
				$S_0 =80$&0.0025& 1.1291& 1.1292& 1.1293 &100000 &1.14$\pm$0.03 &1.14$\pm$0.03
				&1.13$\pm$0.03&1.13$\pm$0.03 &\\
				&0.00125& 1.1291& 1.1292 & &200000 &1.13$\pm$0.02 &1.14$\pm$0.02  &1.11$\pm$0.02&1.12$\pm$0.02&\\
				\hline
				& 0.01& 3.3331& 3.3312 & & 10000&3.27$\pm$0.17 &3.27$\pm$0.17  &3.19$\pm$0.16&3.22$\pm$0.16&\\\
				&0.005& 3.3315& 3.3301 & &50000 &3.32$\pm$0.08 &3.40$\pm$0.08  &3.24$\pm$0.07&3.26$\pm$0.0&\\
				$S_0 =90$&0.0025& 3.3311& 3.3298& 3.3284  &100000 &3.34$\pm$0.05 &3.34$\pm$0.05  &3.32$\pm$0.05&3.33$\pm$0.05&\\
				&0.00125& 3.3310& 3.3297 & &200000 &3.32$\pm$0.04 &3.35$\pm$0.04  &3.28$\pm$0.04&3.31$\pm$0.04&\\
				\hline
				& 0.01& 7.5245& 7.5239 & & 10000&7.46$\pm$0.25 &7.46$\pm$0.25  &7.37$\pm$0.24&7.36$\pm$0.25&\\
				&0.005& 7.5236& 7.5224 & &50000 &7.53$\pm$0.11 &7.62$\pm$0.11  &7.40$\pm$0.11&7.43$\pm$0.11&\\
				$S_0 =100$&0.0025& 7.5231& 7.5221& 7.5210 &100000 &7.54$\pm$0.08 &7.52$\pm$0.08  &7.53$\pm$0.08&7.52$\pm$0.08&\\
				&0.00125& 7.5230& 7.5220 & &200000 &7.50$\pm$0.06 &7.54$\pm$0.06  &7.46$\pm$0.06&7.50$\pm$0.06&\\
				\hline
				& 0.01& 13.6943& 13.6940 & & 10000&13.69$\pm$0.34 &13.69$\pm$0.34  &13.52$\pm$0.33&13.48$\pm$0.33&\\
				&0.005& 13.6923& 13.6924 & &50000 &13.71$\pm$0.15 &13.81$\pm$0.15
				&13.55$\pm$0.15&13.58$\pm$0.15 &\\
				$S_0 =110$&0.0025& 13.6918& 13.6921& 13.6923 &100000 &13.72$\pm$0.11
				&13.69$\pm$0.11  &13.67$\pm$0.11&13.70$\pm$0.11 &\\
				&0.00125& 13.6917& 13.6920 & &200000 &13.64$\pm$0.08 &13.71$\pm$0.08
				&13.63$\pm$0.07&13.69$\pm$0.08 &\\
				\hline
				& 0.01& 21.3173& 21.3185 & & 10000&21.40$\pm$0.41
				&21.40$\pm$0.41  &21.08$\pm$0.40&21.03$\pm$0.41 &\\
				&0.005& 21.3156& 21.3168 & &50000 &21.35$\pm$0.18 &21.46$\pm$0.19
				&21.17$\pm$0.18&21.21$\pm$0.18 &\\
				$S_0 =120$&0.0025& 21.3152& 21.3164& 21.3174 &100000 &21.36$\pm$0.13
				&21.32$\pm$0.13  &21.29$\pm$0.13&21.33$\pm$0.13 &\\
				&0.00125& 21.3152& 21.3163 & &200000 &21.25$\pm$0.09 &21.33$\pm$0.09
				&21.26$\pm$0.09&21.33$\pm$0.09 &\\
				\hline
		\end{tabular}}
	}
	\quad\quad
	\subtable[]{
		\footnotesize
		{\begin{tabular} {@{}c|lrr|c|rrrrrc@{}} \toprule $\rho=0.5$ & $\Delta y$ &HTFDa
				&HTFDb & CF  &$N_{\mathrm{MC}} $&HMCa  &HMCb &AMCa  &AMCb\\
				\hline
				& 0.01 & 1.4732 & 1.4751 & & 10000&1.42$\pm$0.12
				&1.40$\pm$0.12  &1.37$\pm$0.12&1.35$\pm$0.12 &\\
				&0.005 &1.4724 &1.4744 &  &50000 &1.49$\pm$0.06 &1.47$\pm$0.05
				&1.40$\pm$0.05&1.42$\pm$0.05 &\\
				$S_0 =80$ &0.0025 &1.4723 &1.4742 &1.4760 &100000 &1.48$\pm$0.04 &1.46$\pm$0.04  &1.46$\pm$0.04&1.49$\pm$0.04&\\
				&0.00125 &1.4722 &1.4741 & &200000 &1.47$\pm$0.03 &1.48$\pm$0.03  &1.48$\pm$0.03&1.48$\pm$0.03&\\
				\hline
				& 0.01& 3.6849& 3.6859 & & 10000&3.63$\pm$0.19 &3.63$\pm$0.19
				&3.48$\pm$0.19&3.49$\pm$0.19 &\\
				&0.005 &3.6836 &3.6849 & &50000 &3.70$\pm$0.09 &3.70$\pm$0.09  &3.57$\pm$0.09&3.60$\pm$0.09&\\
				$S_0 =90$&0.0025 &3.6832 &3.6847 &3.6862 &100000 &3.67$\pm$0.06 &3.67$\pm$0.06 &3.66$\pm$0.06&3.71$\pm$0.06 &\\
				&0.00125 &3.6832 &3.6847 & &200000 &3.66$\pm$0.04 &3.70$\pm$0.04  &3.69$\pm$0.04&3.68$\pm$0.04&\\
				\hline
				& 0.01& 7.6247 &7.6245 & & 10000&7.58$\pm$0.28
				&7.58$\pm$0.28  &7.35$\pm$0.28&7.36$\pm$0.27 &\\
				&0.005 &7.6238 &7.6232 & &50000 &7.66$\pm$0.13 &7.65$\pm$0.13
				&7.47$\pm$0.12&7.52$\pm$0.12 &\\
				$S_0 =100$&0.0025& 7.6234& 7.6229 &7.6223 &100000 &7.61$\pm$0.09 &7.59$\pm$0.09  &7.58$\pm$0.09&7.66$\pm$0.09&\\
				&0.00125& 7.6233 &7.6228 & &200000 &7.58$\pm$0.06 &7.64$\pm$0.06  &7.62$\pm$0.06&7.61$\pm$0.06 &\\
				\hline
				& 0.01& 13.4863& 13.4835 & & 10000&13.48$\pm$0.36
				&13.48$\pm$0.36  &13.21$\pm$0.36&13.19$\pm$0.36 &\\
				&0.005& 13.4842& 13.4818 & &50000 &13.55$\pm$0.17 &13.49$\pm$0.16
				&13.27$\pm$0.16&13.35$\pm$0.16 &\\
				$S_0 =110$&0.0025& 13.4837 &13.4814 &13.4791 &100000 &13.47$\pm$0.12
				&13.41$\pm$0.12  &13.44$\pm$0.12&13.54$\pm$0.12 &\\
				&0.00125 &13.4836 &13.4813 & &200000 &13.42$\pm$0.08 &13.49$\pm$0.08
				&13.47$\pm$0.08&13.48$\pm$0.08 &\\
				\hline
				& 0.01 &20.9678 &20.9661 & & 10000&21.04$\pm$0.44
				&21.04$\pm$0.44  &20.67$\pm$0.44&20.64$\pm$0.43 &\\
				&0.005 &20.9659 &20.9642 & &50000 &21.05$\pm$0.20 &20.98$\pm$0.20
				&20.71$\pm$0.20&20.81$\pm$0.20 &\\
				$S_0 =120$&0.0025 &20.9655 &20.9636 &20.9616 &100000 &20.96$\pm$0.14
				&20.87$\pm$0.14  &20.92$\pm$0.14&21.04$\pm$0.14 &\\
				&0.00125& 20.9654 &20.9635 & &200000 &20.88$\pm$0.10 &20.96$\pm$0.10
				&20.97$\pm$0.10&20.98$\pm$0.10 &\\
				\hline
		\end{tabular}}
	}
	\caption{\em \small{Standard Bates model. Prices of European call options. Test parameters: $K=100$, $T=0.5$,
			$r=0.03$, $\eta=0.05$, $V_0=0.04$, $\theta_V=0.04$, $\kappa_V=2$, $\sigma_V=0.4$,
			$\lambda=5$, $\gamma=0$, $\delta=0.1$, $\rho=-0.5, 0.5$.}}
	\label{tab1}
\end{table}

\begin{table}[ht] \centering
	\subtable[]{
		\footnotesize
		{\begin{tabular} {@{}c|lrr|cc|rrrrr@{}} \toprule $\rho=-0.5$ & $\Delta y$ &HTFDa
				&HTFDb & PSOR &MOL &$N_{\mathrm{MC}} $&HMCLSa  &HMCLSb &AMCLSa  &AMCLSb\\
				\hline
				& 0.01& 1.1365& 1.1365 & & & 10000&1.03$\pm$0.08&1.14$\pm$0.09&1.06$\pm$0.09&1.03$\pm$0.09\\
				& 0.005& 1.1356& 1.1358 & & &50000&1.19$\pm$0.04&1.14$\pm$0.04&1.18$\pm$0.04&1.12$\pm$0.04\\
				$S_0=80$& 0.0025& 1.1354& 1.1356& 1.1359& 1.1363 &100000&1.15$\pm$0.03&1.13$\pm$0.03&1.13$\pm$0.03&1.13$\pm$0.03 \\
				& 0.00125& 1.1353& 1.1355 & & &200000&1.14$\pm$0.02&1.14$\pm$0.02&1.14$\pm$0.02&1.14$\pm$0.02\\
				\hline
				& 0.01& 3.3579& 3.3563 & & &10000&3.39$\pm$0.15&3.44$\pm$0.16&3.38$\pm$0.15&3.48$\pm$0.16 \\
				& 0.005& 3.3564& 3.3551 & & &50000&3.46$\pm$0.07&3.33$\pm$0.07&3.46$\pm$0.07&3.32$\pm$0.07 \\
				$S_0=90$& 0.0025& 3.3560& 3.3548& 3.3532& 3.3530 &100000&3.35$\pm$0.05&3.35$\pm$0.05&3.33$\pm$0.05&3.36$\pm$0.05 \\
				& 0.00125& 3.3559& 3.3547 & & &200000&3.35$\pm$0.03&3.33$\pm$0.03&3.35$\pm$0.03&3.34$\pm$0.03 \\
				\hline
				& 0.01& 7.6010& 7.6006 & & &10000&7.68$\pm$0.23&7.88$\pm$0.24&7.63$\pm$0.23&7.80$\pm$0.24 \\
				& 0.005& 7.6001& 7.5992 &  & &50000&7.75$\pm$0.11&7.59$\pm$0.10&7.76$\pm$0.10&7.53$\pm$0.10\\
				$S_0=100$& 0.0025& 7.5997& 7.5989& 7.5970& 7.5959 &100000&7.56$\pm$0.07&7.61$\pm$0.07&7.56$\pm$0.07&7.61$\pm$0.07 \\
				& 0.00125& 7.5996& 7.5989 &  & &200000&7.58$\pm$0.05&7.55$\pm$0.05&7.58$\pm$0.05&7.57$\pm$0.05\\
				\hline
				& 0.01& 13.8853& 13.8854 & & &10000&13.90$\pm$0.29&14.28$\pm$0.30&13.84$\pm$0.29&14.10$\pm$0.29 \\
				& 0.005& 13.8836& 13.8842 & & &50000&14.05$\pm$0.13&13.89$\pm$0.12&14.07$\pm$0.13&13.86$\pm$0.12 \\
				$S_0=110$& 0.0025& 13.8832& 13.8839& 13.8830& 13.8827 &100000&13.80$\pm$0.09&13.91$\pm$0.09&13.84$\pm$0.09&13.89$\pm$0.09\\
				& 0.00125& 13.8831& 13.8838 & & &200000&13.86$\pm$0.06&13.84$\pm$0.06&13.87$\pm$0.06&13.83$\pm$0.06 \\
				\hline
				& 0.01& 21.7180& 21.7199 & & &10000&21.83$\pm$0.34&22.07$\pm$0.33&21.71$\pm$0.30&22.04$\pm$0.34\\
				& 0.005& 21.7168& 21.7187 & & &50000&21.91$\pm$0.15&21.76$\pm$0.13&21.90$\pm$0.15&21.72$\pm$0.13 \\
				$S_0=120$& 0.0025& 21.7166& 21.7184 & 21.7186& 21.7191 &100000&21.59$\pm$0.10&21.78$\pm$0.10&21.64$\pm$0.10&21.72$\pm$0.10\\
				& 0.00125& 21.7165& 21.7183 & & &200000&21.68$\pm$0.07&21.65$\pm$0.07&21.68$\pm$0.07&21.67$\pm$0.07\\
				\hline
		\end{tabular}}
	}\quad\quad
	\subtable[]{
		\footnotesize
		{\begin{tabular} {@{}c|lrr|cc|rrrrr@{}} \toprule $\rho=0.5$ & $\Delta y$ &HTFDa
				&HTFDb & PSOR &MOL &$N_{\mathrm{MC}} $&HMCLSa  &HMCLSb &AMCLSa  &AMCLSb\\
				\hline
				& 0.01& 1.4817& 1.4837 & & &10000&1.32$\pm$0.11&1.03$\pm$0.09&1.51$\pm$0.13&0.66$\pm$0.08\\
				& 0.005& 1.4809& 1.4830 & & &50000&1.51$\pm$0.05&1.31$\pm$0.05&1.54$\pm$0.05&1.47$\pm$0.05\\
				$S_0=80$& 0.0025& 1.4807& 1.4828& 1.4843& 1.4848 &100000&1.50$\pm$0.04&1.50$\pm$0.04&1.51$\pm$0.04&1.48$\pm$0.04\\
				& 0.00125& 1.4807& 1.4828 & & &200000&1.50$\pm$0.03&1.49$\pm$0.02&1.49$\pm$0.03&1.47$\pm$0.02\\
				\hline
				& 0.01& 3.7134& 3.7148 & & &10000&3.83$\pm$0.19&3.79$\pm$0.17&3.89$\pm$0.19&3.95$\pm$0.19\\
				& 0.005& 3.7121& 3.7139 & & &50000&3.81$\pm$0.08&3.70$\pm$0.08&3.84$\pm$0.08&3.69$\pm$0.08\\
				$S_0=90$& 0.0025& 3.7118& 3.7137& 3.7145& 3.7146 &100000&3.69$\pm$0.06&3.75$\pm$0.06&3.72$\pm$0.06&3.70$\pm$0.06 \\
				& 0.00125& 3.7118& 3.7137 & & &200000&3.70$\pm$0.04&3.71$\pm$0.04&3.72$\pm$0.04&3.70$\pm$0.04\\
				\hline
				& 0.01& 7.7044& 7.7051 & & &10000&7.74$\pm$0.26&7.85$\pm$0.25&7.96$\pm$0.26&7.99$\pm$0.26\\
				& 0.005& 7.7036& 7.7039 & & &50000&7.85$\pm$0.12&7.68$\pm$0.11&7.87$\pm$0.12&7.68$\pm$0.11 \\
				$S_0=100$& 0.0025& 7.7033& 7.7036& 7.7027& 7.7018 &100000&7.66$\pm$0.08&7.75$\pm$0.08&7.65$\pm$0.08&7.73$\pm$0.08\\
				& 0.00125& 7.7032& 7.7036 & & &200000&7.69$\pm$0.06&7.67$\pm$0.05&7.68$\pm$0.06&7.69$\pm$0.05\\
				\hline
				& 0.01& 13.6770& 13.6756 & & &10000&13.57$\pm$0.32&13.98$\pm$0.31&13.88$\pm$0.32&14.12$\pm$0.33\\
				& 0.005& 13.6752& 13.6742 &  & &50000&13.83$\pm$0.14&13.67$\pm$0.13&13.89$\pm$0.14&13.64$\pm$0.13\\
				$S_0=110$& 0.0025& 13.6747& 13.6739& 13.6722& 13.6715 &100000&13.56$\pm$0.09&13.74$\pm$0.10&13.58$\pm$0.10&13.71$\pm$0.10\\
				& 0.00125& 13.6747& 13.6738 & & &200000&13.65$\pm$0.07&13.65$\pm$0.07&13.64$\pm$0.07&13.64$\pm$0.07\\
				\hline
				& 0.01& 21.3668& 21.3671& & &10000&21.45$\pm$0.32&21.60$\pm$0.35&21.39$\pm$0.33&21.84$\pm$0.34 \\
				& 0.005& 21.3655& 21.3658 & & &50000&21.54$\pm$0.15&21.40$\pm$0.14&21.61$\pm$0.16&21.40$\pm$0.13\\
				$S_0=120$& 0.0025& 21.3653& 21.3655& 21.3653& 21.3657 &100000&21.26$\pm$0.10&21.43$\pm$0.10&21.27$\pm$0.10&21.38$\pm$0.10\\
				& 0.00125& 21.3652& 21.3653 & & &200000&21.31$\pm$0.07&21.33$\pm$0.07&21.31$\pm$0.07&21.31$\pm$0.07 \\
				\hline
		\end{tabular}}
	}
	\caption{\em \small{Standard Bates model. Prices of American call options. Test parameters: $K=100$, $T=0.5$,
			$r=0.03$, $\eta=0.05$, $V_0=0.04$, $\theta_V=0.04$, $\kappa_V=2$, $\sigma_V=0.4$,
			$\lambda=5$, $\gamma=0$, $\delta=0.1$, $\rho=-0.5,0.5$.}}
	\label{tab2}
\end{table}

\begin{table} [ht] \centering
	\footnotesize
	{\begin{tabular} {@{}clcc|rcccc|cc@{}} \toprule & $\Delta y$ &HTFDa &HTDFb
			&$N_{\mathrm{MC}}$&HMCa &HMCb &AMCa &AMCb &CF\\
			\hline
			& 0.01  &0.09 &0.34 &10000 &0.007 &0.16 &0.16 &0.30 & &\\
			& 0.005 &0.18 &0.72  &50000  &0.36 &0.72  &0.79 &1.51 & &\\
			& 0.0025  &0.46 &1.62  &100000 &0.71 &1.44  &1.57 &3.12  &0.001 &\\
			& 0.00125  &0.84 &3.53 &200000 &1.45 &2.95 &3.14 &6.17 & &\\
	\end{tabular}}
	\caption{\em \small{Standard Bates model. Computational times (in seconds) for European call
			options in Table \ref{tab1} for $S_0=100$, $\rho=-0.5.$}}
	\label{tab31}
\end{table}

\begin{table} [ht] \centering
	\footnotesize
	{\begin{tabular} {@{}clcc|rccccc@{}} \toprule & $\Delta y$ &HTFDa &HTDFb
			&$N_{\mathrm{MC}}$&HMCLSa &HMCLSb &AMCLSa &AMCLSb\\
			\hline
			& 0.01  &0.10 &0.37 &10000 &0.09 &0.23 &0.20 &0.45 &\\
			& 0.005 &0.19 &0.77  &50000  &0.47 &1.11  &1.01 &2.25 &\\
			& 0.0025  &0.48 &1.77  &100000 &1.07 &2.25  &2.01 &4.57  &\\
			& 0.00125  &0.95 &3.61 &200000 &1.94 &4.55 &4.05 &8.98 &\\
	\end{tabular}}
	\caption{\em \small{Standard Bates model. Computational times (in seconds) for American call
			options in Table \ref{tab2} for $S_0=100$, $\rho=-0.5.$}}
	\label{tab32}
\end{table}

\begin{table} [ht] \centering
	\footnotesize
	{\begin{tabular} {@{}cccccc@{}}\toprule $N$ &$S_0 =80$
			& $S_0 =90$ & $S_0 =100$ & $S_0 =110$ & $S_0 =120$ \\
			\hline
			200 &1.919250  &1.961063 &1.894156  &2.299666 &2.109026\\
			400 &2.172836 &2.209762 &2.556021  &1.673541 &1.996332\\
			800 &1.544849 & 1.851932 &1.463712  &2.935697 &2.106880\\
			\hline
	\end{tabular}}
	\caption{\em \small{Standard Bates model. HTFDb-ratio \eqref{ratio} for the price of
			American call options as the starting point $S_0$ varies with
			fixed space step $\Delta y=0.0025$. Test parameters: $T=0.5$,
			$r=0.03$, $\eta=0.05$, $V_0=0.04$, $\theta=0.04$, $\kappa=2$, $\sigma=0.4$,
			$\lambda=5$, $\gamma=0$, $\delta=0.1$, $\rho=-0.5$.}}
	\label{tab4ratio}
\end{table}

\clearpage

\subsubsection{Options with large maturity in the
	standard Bates model}\label{andersen}
In order to verify the robustness of the proposed algorithms we consider
experiments when the Feller condition $2\kappa_V \theta_V\geq\sigma_V^2$  is not fulfilled for the CIR volatility process. We additionally stress our tests by considering  large maturities. For this purpose we consider the parameters from Chiarella \emph{et al.} \cite{ckmz} already used in Section \ref{sect-num-bates} with
$\rho=-0.5$, except for the maturity and the vol-vol, which are modified as follows: $T=5$ and $\sigma_V=0.7$ respectively.

Table \ref{tab4} reports European call option
prices, which are compared with the true values (\textbf{CF}). In Table \ref{tab5} we provide results for
American call option prices.
The settings for the experiments \textbf{HTFDa-b}, \textbf{HMCa-b} and
\textbf{AMCa-b} are the same as described at the beginning of Section \ref{sect-num-bates}.
The settings for the experiments in the American case
\textbf{HMCLSa-b} and \textbf{AMCLSa-b} are changed
\begin{itemize}
	\item[]
	\textbf{HMCLSa} and \textbf{AMCLSa}: $20$ exercise dates, $N_t=100$ and $N_{\mathrm{MC}}=10000, 50000, 100000, 200000$;
	\item[]
	\textbf{HMCLSb} and \textbf{AMCLSb}: $40$ exercise dates, $N_t=200$ and $N_{\mathrm{MC}}=10000, 50000, 100000$, $200000$.
\end{itemize}
In the American case the benchmark values \textbf{B-AMC} are obtained by the Longstaff-Schwartz \cite{ls} Monte Carlo algorithm with $300$
exercise dates, combined with the accurate
third-order Alfonsi method with $3000$ discretization time steps and 1
million iterations.

The numerical results suggest that large maturities bring to a slight loss of accuracy for \textbf{HTFD} and \textbf{HMC}, even if both methods
provide a satisfactory approximation of the true option prices, being in turn mostly compatible with the results from the Alfonsi Monte Carlo method. It is
worth noticing that for long maturity $T=5$ we have developed
experiments with the same number of steps both in time ($N_t$) and
space step ($\Delta y$) as for $T=0.5$. So, the numerical experiments are not slower, and it is clear that one could achieve a better accuracy for larger values of $N_t$.
\begin{table}[ht]
	\centering
	\footnotesize
	{\begin{tabular} {@{}c|lrr|c|rrrrrc@{}} \toprule $\rho=-0.5$ & $\Delta y$
			&HTFDa  &HTFDb & CF &$N_{\mathrm{MC}} $&HMCa  &HMCb  &AMCa  &AMCb\\
			\hline
			& 0.01&9.0085 &8.9457  & & 10000 &9.21$\pm$0.55 &9.09$\pm$0.55
			&8.69$\pm$0.53&8.56$\pm$0.51 &\\
			&0.0050 &9.0032 &8.9405  &  &50000 &9.13$\pm$0.25 &8.92$\pm$0.24
			&8.81$\pm$0.24&9.04$\pm$0.24 &\\
			$S_0=80$&0.0025 &9.0020 &8.9392  &8.9262  &100000 &9.01$\pm$0.17
			&8.81$\pm$0.17  &8.92$\pm$0.17&8.88$\pm$0.17  &\\\
			&0.00125 &9.0016 &8.9389  &  &200000 &8.99$\pm$0.12 &8.92$\pm$0.12  &8.95$\pm$0.12&8.90$\pm$0.12 &\\
			\hline
			& 0.01&12.7405 &12.6520   & & 10000 &12.95$\pm$0.67
			&12.95$\pm$0.67  &12.29$\pm$0.65&12.15$\pm$0.6 &\\
			&0.0050 &12.7342 &12.6458  & &50000 &12.87$\pm$0.30 &12.64$\pm$0.29
			&12.49$\pm$0.29&12.76$\pm$0.3 &\\
			$S_0=90 $&0.0025 &12.7327 &12.6442  &12.6257  &100000 &12.72$\pm$0.21
			&12.50$\pm$0.21  &12.63$\pm$0.21&12.58$\pm$0.21 &\\
			&0.00125 &12.7323 &12.6438  &  & 200000 &12.71$\pm$0.15 &12.61$\pm$0.15
			&12.66$\pm$0.15&12.61$\pm$0.15 &\\
			\hline
			& 0.01&17.0324 &16.9176  & &10000  &17.24$\pm$0.80
			&17.24$\pm$0.80  &16.43$\pm$0.77&16.29$\pm$0.75 &\\
			&0.0050 &17.0254 &16.9106  &  &50000 &17.18$\pm$0.36 &16.91$\pm$0.35
			&16.73$\pm$0.35&17.03$\pm$0.35 &\\
			$S_0=100 $&0.0025 &17.0237 &16.9089 &16.8855  &100000  &17.00$\pm$0.25
			&16.74$\pm$0.25  &16.91$\pm$0.25&16.84$\pm$0.25 &\\
			&0.00125 &17.0232 &16.9084  & &200000 &16.99$\pm$0.18 &16.86$\pm$0.18
			&16.94$\pm$0.18&16.88$\pm$0.18 &\\
			\hline
			& 0.01&21.8149 &21.6741   & &10000 &22.04$\pm$0.93
			&22.04$\pm$0.93  &21.06$\pm$0.93&20.91$\pm$0.88 &\\
			&0.0050 &21.8067 &21.6659  & &50000 &21.96$\pm$0.42 &21.67$\pm$0.41 &21.43$\pm$0.41&21.82$\pm$0.41 &\\
			$S_0=110 $&0.0025 &21.8047 &21.6639  &21.6364 &100000 &21.76$\pm$0.29
			&21.47$\pm$0.29  &21.69$\pm$0.29&21.59$\pm$0.29 &\\
			&0.00125 &21.8042 &21.6634  & &200000 &21.76$\pm$0.21 &21.59$\pm$0.20
			&21.70$\pm$0.20&21.63$\pm$0.20 &\\
			\hline
			& 0.01&27.0196 &26.8539  & &10000 &27.26$\pm$1.05
			&27.26$\pm$1.05  &26.12$\pm$1.03&25.94$\pm$1.01 &\\
			&0.0050 &27.0108 &26.8452  & &50000 &27.17$\pm$0.47 &26.86$\pm$0.46
			&26.56$\pm$0.46&27.02$\pm$0.47 &\\
			$S_0=120 $&0.0025 &27.0086 &26.8430  &26.8121 &100000 &26.94$\pm$0.33
			&26.63$\pm$0.33  &26.89$\pm$0.33&26.78$\pm$0.33 &\\
			&0.00125 &27.0081 &26.8425  &  &200000 &26.95$\pm$0.23 &26.75$\pm$0.23
			&26.89$\pm$0.23&26.81$\pm$0.23 &\\
			\hline
		\end{tabular}
	}
	\caption{\em \small{Standard Bates model. Prices of European call options. Test parameters: $K=100$, $T=5$,
			$r=0.03$, $\eta=0.05$, $V_0=0.04$, $\theta_V=0.04$, $\kappa_V=2$, $\sigma_V=0.7$,
			$\lambda=5$, $\gamma=0$, $\delta=0.1$, $\rho=-0.5$. Case $2\kappa_V\theta_V<\sigma^2_V$.}}
	\label{tab4}
\end{table}

\begin{table}[ht]\centering
	\footnotesize
	{\begin{tabular} {@{}c|lrr|c|rrrrr@{}} \toprule $\rho=-0.5$ & $\Delta y$ &HTFDa
			&HTFDb &B-AMC &$N_{\mathrm{MC}} $&HMCLSa  &HMCLSb &AMCLSa  &AMCLSb\\
			\hline
			& 0.01&9.8335 &9.7978  & &10000&10.15$\pm$0.46&10.20$\pm$0.46&10.47$\pm$0.47&9.80$\pm$0.42\\
			&0.0050 &9.8283 &9.7927  & &50000&9.93$\pm$0.20&9.86$\pm$0.20&9.89$\pm$0.19&9.78$\pm$0.19 \\
			$S_0=80$&0.0025 &9.8271 &9.7914  &9.7907$\pm$ 0.04 &100000&9.76$\pm$0.14&9.69$\pm$0.13&9.74$\pm$0.14&9.76$\pm$0.13\\
			&0.00125 &9.8267 &9.7911  & &200000&9.79$\pm$0.10&9.70$\pm$0.09&9.73$\pm$0.10&9.72$\pm$0.09 \\
			\hline
			& 0.01&14.0801 &14.0318  & &10000&14.58$\pm$0.56&14.46$\pm$0.55&14.94$\pm$0.58&14.08$\pm$0.51 \\
			&0.0050 &14.0741 &14.0258  & &50000&14.13$\pm$0.24&14.14$\pm$0.24&14.19$\pm$0.23&14.12$\pm$0.23\\
			$S_0=90 $&0.0025 &14.0726 &14.0244  &14.0030$\pm$ 0.05 &100000&13.98$\pm$0.16&13.87$\pm$0.16&13.94$\pm$0.16&13.89$\pm$0.16 \\
			&0.00125 &14.0722 &14.0240  & &200000&13.93$\pm$0.12&13.91$\pm$0.11&13.94$\pm$0.12&13.96$\pm$0.11 \\
			\hline
			& 0.01&19.0658 &19.0075  & &10000&19.59$\pm$0.66&19.44$\pm$0.63&19.88$\pm$0.66&19.13$\pm$0.59 \\
			&0.0050 &19.0594 &19.0011  & &50000&19.10$\pm$0.27&19.06$\pm$0.27&19.26$\pm$0.26&19.01$\pm$0.26\\
			$S_0=100 $&0.0025 &19.0578 &18.9995  &18.9632$\pm$ 0.05 &100000&18.92$\pm$0.19&18.88$\pm$0.18&18.85$\pm$0.19&18.90$\pm$0.18 \\
			&0.00125 &19.0574 &18.9991  & &200000&18.80$\pm$0.13&18.84$\pm$0.13&18.85$\pm$0.13&18.92$\pm$0.13 \\
			\hline
			& 0.01&24.7434 &24.6788  & &10000&25.02$\pm$0.74&24.84$\pm$0.72&25.32$\pm$0.72&24.78$\pm$0.67\\
			&0.0050 &24.7364 &24.6719  & &50000&24.79$\pm$0.30&24.57$\pm$0.29&24.94$\pm$0.29&24.72$\pm$0.29\\
			$S_0=110 $&0.0025 &24.7347 &24.6701  &24.6289$\pm$ 0.06 &100000&24.53$\pm$0.21&24.47$\pm$0.20&24.50$\pm$0.21&24.51$\pm$0.20\\
			&0.00125 &24.7343 &24.6697  & &200000&24.42$\pm$0.14&24.45$\pm$0.14&24.50$\pm$0.15&24.53$\pm$0.14\\
			\hline
			& 0.01&31.0646 &30.9983  & &10000&30.88$\pm$0.74&31.15$\pm$0.75&31.18$\pm$0.74&31.04$\pm$0.71\\
			&0.0050 &31.0577 &30.9914  & &50000&31.10$\pm$0.32&30.94$\pm$0.31&31.32$\pm$0.32&30.98$\pm$0.32\\
			$S_0=120 $&0.0025 &31.0559 &30.9896  &30.9052$\pm$0.07 &100000&30.89$\pm$0.23&30.72$\pm$0.22&30.70$\pm$0.22&30.72$\pm$0.22\\
			&0.00125 &31.0555 &30.9892  & &200000&30.72$\pm$0.16&30.73$\pm$0.16&30.77$\pm$0.16&30.89$\pm$0.15\\
			\hline
		\end{tabular}
	}
	\caption{\em \small{Standard Bates model. Prices of American call options. Test parameters: $K=100$, $T=5$,
			$r=0.03$, $\eta=0.05$, $V_0=0.04$, $\theta_V=0.04$, $\kappa_V=2$, $\sigma_V=0.7$,
			$\lambda=5$, $\gamma=0$, $\delta=0.1$, $\rho=-0.5$. Case $2\kappa_V\theta_V<\sigma^2_V$.}}
	\label{tab5}
\end{table}


\subsubsection{Bates model with stochastic interest rate}\label{sect-num-bateshw}
We consider now the case of Bates model associated with the Vasiceck model for the stochastic interest rate.
For the Bates model we consider the parameters from Chiarella \emph{et al.}
\cite{ckmz} already used in Section \ref{sect-num-bateshw}. Moreover, for the interest rate parameter we fix the following parameters:
initial interest rate $r_0=0.03$, speed of mean-reversion   $\kappa_r=1$, interest rate volatility  $\sigma_r=0.2$;
	time-varying long-term mean $\theta_r(t)$ fitting the theoretical bond prices to the yield curve observed on the market, here set as
	$P_r(0,T)=e^{-0.03T}$. We study the cases $\rho_1=\rho_{SV}=-0.5$ and $\rho_2=\rho_{Sr}=-0.5, 0.5$.
No correlation is assumed to exist between $r$ and $V$.
We consider the mesh grid $\Delta
y=0.02$, $0.01$, $0.005$, $0.0025$,
the case $\Delta y=0.00125$ being removed because it requires huge computational times.
The numerical results are labeled \textbf{HTFDa-b},
\textbf{HMCa-b}, \textbf{AMCa-b}, \textbf{HMCLSa-b}, \textbf{AMCLSa-b}, their settings being given at the beginning of Section \ref{sect-num-bates}.

When the interest rate is assumed to be stochastic, no references are available in the literature. Therefore, we propose benchmark values obtained by using a Monte Carlo method in which the CIR paths are simulated  through the accurate
third-order Alfonsi \cite{al} discretization scheme and the interest rate paths are generated by an exact scheme. For these benchmark values, called \textbf{B-AMC}, the number of Monte Carlo iterations and of the discretization time steps are set as $N_{\mathrm{MC}}=10^6$  and
$N_t=300$ respectively. In the American case, \textbf{B-AMC} is evaluated through the Longstaff-Schwartz \cite{ls} algorithm  with $20$
exercise dates. All Monte Carlo results report the $95\% $ confidence
intervals.

European and American call option prices are given in tables \ref{tab6} and \ref{tab7} respectively. Tables \ref{tab81} and \ref{tab82} refer to the computational time cost (in seconds) of the different algorithms in the European Call case and American Call case respectively.
The numerical results confirm the good numerical behavior of \textbf{HTFD}  and
\textbf{HMC} in the Bates-Hull-White model as well.

\begin{table}[ht]
	\centering
	\subtable[]{
		\footnotesize
		{\begin{tabular} {@{}c|lrr|c|rrrrrr@{}} 
				\toprule $\rho_{Sr}=-0.5$ & $\Delta y$
				&HTFDa  &HTFDb & B-AMC &$N_{\mathrm{MC}} $&HMCa  &HMCb  &AMCa  &AMCb\\
				\hline
				& 0.02&1.0169 &1.0079  & &  10000 &1.00$\pm$0.09
				&0.96$\pm$0.09  &1.00$\pm$0.09&1.06$\pm$0.10 & \\
				&0.01 &1.0201 &1.0188  & &50000 &1.02$\pm$0.04 &0.97$\pm$0.04
				&0.98$\pm$0.04&1.01$\pm$0.04 &\\
				$S_0=80$&0.0050 &1.0199 &1.0194  &1.0153$\pm0.01$ &100000 &1.00$\pm$0.03
				&1.00$\pm$0.03  &1.01$\pm$0.03&1.03$\pm$0.03 &\\
				&0.0025 &1.0197 &1.0193  & &200000 &1.01$\pm$0.02 &1.01$\pm$0.02
				&1.02$\pm$0.02&1.00$\pm$0.02 &\\
				\hline
				& 0.01&3.1172 &3.1032  & & 10000&3.05$\pm$0.16 &3.05$\pm$0.16
				&3.07$\pm$0.16&3.14$\pm$0.17 &\\
				&0.01 &3.1186 &3.1137  & &50000 &3.10$\pm$0.07 &3.03$\pm$0.07
				&3.02$\pm$0.07&3.09$\pm$0.07 &\\
				$S_0=90 $&0.0050 &3.1174 &3.1135  &3.1008$\pm0.02$
				&100000 &3.07$\pm$0.05 &3.08$\pm$0.05  &3.09$\pm$0.05&3.14$\pm$0.05 &\\
				&0.0025 &3.1174 &3.1136  & &200000 &3.09$\pm$0.04 &3.10$\pm$0.04  &3.11$\pm$0.04&3.08$\pm$0.04 &\\
				\hline
				& 0.02&7.2528 &7.2472  & & 10000&7.17$\pm$0.24
				&7.17$\pm$0.24  &7.20$\pm$0.24&7.24$\pm$0.25 &\\
				&0.01 &7.2528 &7.2479  & &50000 &7.21$\pm$0.11 &7.18$\pm$0.11
				&7.12$\pm$0.11&7.21$\pm$0.11 &\\
				$S_0=100 $&0.0050 &7.2528 &7.2480  &7.2315$\pm 0.02$ &100000 &7.18$\pm$0.08
				&7.24$\pm$0.08  &7.20$\pm$0.08&7.27$\pm$0.08 &\\
				&0.0025 &7.2528 &7.2480  & &200000 &7.22$\pm$0.05 &7.25$\pm$0.05 &7.24$\pm$0.05&7.20$\pm$0.05 &\\
				\hline
				& 0.02&13.4553 &13.4565  & & 10000&13.30$\pm$0.32
				&13.30$\pm$0.32  &13.41$\pm$0.33&13.39$\pm$0.33 &\\
				&0.01 &13.4465 &13.4440  & &50000 &13.37$\pm$0.15 &13.40$\pm$0.15
				&13.27$\pm$0.15&13.38$\pm$0.15 &\\
				$S_0=110 $&0.0050 &13.4435 &13.4407  &13.4256$\pm 0.03$ &100000 &13.35$\pm$0.10
				&13.46$\pm$0.10  &13.38$\pm$0.10&13.48$\pm$0.10 &\\
				&0.0025 &13.4432 &13.4404  & &200000 &13.40$\pm$0.07 &13.47$\pm$0.07
				&13.43$\pm$0.07&13.39$\pm$0.07 &\\
				\hline
				& 0.02&21.1320 &21.1356  & & 10000&20.89$\pm$0.40
				&20.89$\pm$0.40  &21.08$\pm$0.40&20.99$\pm$0.41  &\\
				&0.01 &21.1243 &21.1239  & &50000 &21.03$\pm$0.18 &21.09$\pm$0.18
				&20.92$\pm$0.18&21.03$\pm$0.18 &\\
				$S_0=120 $&0.0050 &21.1222 &21.1214  &21.1070$\pm 0.04$ &100000 &21.01$\pm$0.13
				&21.17$\pm$0.13  &21.04$\pm$0.13&21.17$\pm$0.13 &\\
				&0.0025 &21.1215 &21.1207  & &200000 &21.06$\pm$0.09 &21.16$\pm$0.09 &21.12$\pm$0.09&21.06$\pm$0.09 &\\
				\hline
		\end{tabular}}
	}
	\quad\quad
	\subtable[]{
		\footnotesize
		{\begin{tabular} {@{}c|lrr|c|rrrrrr@{}} \toprule $\rho_{Sr}=0.5$ & $\Delta y$ &HTFDa
				&HTFDb & B-AMC  &$N_{\mathrm{MC}} $&HMCa  &HMCb  &AMCa  &AMCb\\
				\hline
				& 0.02&1.3459 &1.3379  & & 10000&1.29$\pm$0.11 &1.28$\pm$0.11
				&1.32$\pm$0.10&1.41$\pm$0.11 &\\
				&0.01 &1.3482 &1.3471  & &50000 &1.34$\pm$0.05 &1.30$\pm$0.05
				&1.32$\pm$0.05&1.35$\pm$0.05  &\\
				$S_0=80$&0.0050 &1.3479 &1.3475  &1.3446$\pm$0.01 &100000 &1.32$\pm$0.03
				&1.31$\pm$0.03  &1.34$\pm$0.03&1.34$\pm$0.03 &\\
				&0.0025 &1.3477 &1.3473  & &200000 &1.33$\pm$0.02 &1.34$\pm$0.02 &1.35$\pm$0.02&1.32$\pm$0.02 &\\
				\hline
				& 0.01&3.7320 &3.7233  & & 10000&3.62$\pm$0.18 &3.62$\pm$0.18
				&3.64$\pm$0.18&3.76$\pm$0.19  &\\
				&0.01 &3.7323 &3.7304  & &50000 &3.69$\pm$0.08 &3.65$\pm$0.08 &3.64$\pm$0.18&3.76$\pm$0.19 &\\
				$S_0=90 $&0.0050 &3.7311 &3.7298  &3.7263$\pm 0.02$ &100000 &3.66$\pm$0.06
				&3.68$\pm$0.06  &3.71$\pm$0.06&3.73$\pm$0.06 &\\
				&0.0025 &3.7311 &3.7299  & &200000 &3.69$\pm$0.04 &3.72$\pm$0.04 &3.73$\pm$0.04&3.68$\pm$0.04  &\\
				\hline
				& 0.02&8.0100 &8.0073  & & 10000&7.83$\pm$0.26
				&7.83$\pm$0.26  &7.82$\pm$0.26&8.00$\pm$0.27 &\\
				&0.01 &8.0112 &8.0102  & &50000 &7.92$\pm$0.12 &7.93$\pm$0.12
				&7.93$\pm$0.12&7.97$\pm$0.12 &\\
				$S_0=100 $&0.0050 &8.0114 &8.0107  &8.0069$\pm 0.03$ &100000 &7.91$\pm$0.08 &7.97$\pm$0.08 &7.99$\pm$0.08&8.02$\pm$0.08 &\\
				&0.0025 &8.0114 &8.0107  & &200000 &7.95$\pm$0.06 &8.02$\pm$0.06
				&8.00$\pm$0.06&7.95$\pm$0.06 &\\
				\hline
				& 0.02&14.1482 &14.1505  & & 10000&13.89$\pm$0.35
				&13.89$\pm$0.35  &13.88$\pm$0.35&14.07$\pm$0.36 &\\
				&0.01 &14.1413 &14.1414  & &50000 &14.01$\pm$0.16 &14.05$\pm$0.16 &14.03$\pm$0.16&14.09$\pm$0.16 &\\
				$S_0=110 $&0.0050 &14.1388 &14.1388  &14.1323$\pm 0.03$ &100000 &14.01$\pm$0.11
				&14.10$\pm$0.11  &14.12$\pm$0.11&14.14$\pm$0.11 &\\
				&0.0025 &14.1386 &14.1386  & &200000 &14.06$\pm$0.08 &14.17$\pm$0.08
				&14.13$\pm$0.08&14.07$\pm$0.08 &\\
				\hline
				& 0.02&21.6737 &21.6772  & & 10000&21.37$\pm$0.42
				&21.37$\pm$0.42  &21.35$\pm$0.42&21.51$\pm$0.43 &\\
				&0.01 &21.6670 &21.6674  & &50000 &21.50$\pm$0.19 &21.55$\pm$0.19 &21.52$\pm$0.19&21.60$\pm$0.19 &\\
				$S_0=120 $&0.0050 &21.6651 &21.6653  &21.6501$\pm 0.04$ &100000 &21.52$\pm$0.13
				&21.63$\pm$0.13  &21.64$\pm$0.13&21.68$\pm$0.14 &\\
				&0.0025 &21.6645 &21.6646  & &200000 &21.57$\pm$0.10 &21.71$\pm$0.10  &21.65$\pm$0.10&21.58$\pm$0.09&\\
				\hline
		\end{tabular}}
	}
	\caption{\em \small{Bates-Hull-White model. Prices of European call options. Test parameters: $K=100$, $T=0.5$,
			$\eta=0.05$, , $r_0=0.03$, $\kappa_r=1$, $\sigma_r=0.2$, $V_0=0.04$, $\theta_V=0.04$, $\kappa_V=2$, $\sigma_V=0.4$,
			$\lambda=5$, $\gamma=0$, $\delta=0.1$,
			$\rho_{SV}=-0.5$,$\rho_{Sr}=-0.5, 0.5$.}}
	\label{tab6}
\end{table}

\begin{table}[ht] \centering
	\subtable[]{
		\footnotesize
		{\begin{tabular} {@{}c|lrr|c|rrrrr@{}} \toprule $\rho_{Sr}=-0.5$ & $\Delta y$ &HTFDa
				&HTFDb & B-AMC &$N_{\mathrm{MC}} $&HMCLSa  &HMCLSb &AMCLSa  &AMCLSb\\
				\hline
				& 0.02&1.0561 &1.0470  & &10000&0.76$\pm$0.07&0.56$\pm$0.06&0.95$\pm$0.08&0.82$\pm$0.08 \\
				&0.01 &1.0598 &1.0588  & &50000&1.08$\pm$0.04&0.91$\pm$0.04&1.01$\pm$0.04&0.96$\pm$0.04 \\
				$S_0=80$&0.0050 &1.0597 &1.0596  &1.0544$\pm0.01$ &100000&1.07$\pm$0.03&1.03$\pm$0.03&1.07$\pm$0.03&1.04$\pm$0.03
				\\
				&0.0025 &1.0596 &1.0595  & &200000&1.05$\pm$0.02&1.04$\pm$0.02&1.07$\pm$0.02&1.05$\pm$0.02\\
				\hline
				& 0.01&3.2511 &3.2364  & &10000&3.28$\pm$0.15&3.39$\pm$0.16&3.35$\pm$0.16&3.07$\pm$0.15 \\
				&0.01 &3.2537 &3.2493  & &50000&3.33$\pm$0.07&3.21$\pm$0.07&3.25$\pm$0.07&3.30$\pm$0.07\\
				$S_0=90 $&0.0050 &3.2528 &3.2494  &3.2273$\pm 0.01$ &100000&3.23$\pm$0.05&3.24$\pm$0.05&3.27$\pm$0.05&3.25$\pm$0.05 \\
				&0.0025 &3.2528 &3.2495  & &200000&3.22$\pm$0.03&3.23$\pm$0.03&3.25$\pm$0.03&3.24$\pm$0.03 \\
				\hline
				& 0.02&7.6012 &7.5952  & &10000&7.64$\pm$0.22&7.99$\pm$0.23&7.80$\pm$0.23&7.68$\pm$0.22\\
				&0.01 &7.6020 &7.5976  & &50000&7.72$\pm$0.10&7.58$\pm$0.09&7.61$\pm$0.10&7.65$\pm$0.10\\
				$S_0=100 $&0.0050 &7.6022 &7.5980  &7.5589$\pm 0.02$ &100000&7.54$\pm$0.07&7.62$\pm$0.07&7.61$\pm$0.07&7.54$\pm$0.07 \\
				&0.0025 &7.6022 &7.5980  & &200000&7.54$\pm$0.05&7.54$\pm$0.05&7.56$\pm$0.05&7.60$\pm$0.05\\
				\hline
				& 0.02&14.1510 &14.1524  & &10000&14.22$\pm$0.28&14.61$\pm$0.29&14.35$\pm$0.29&14.07$\pm$0.28\\
				&0.01 &14.1443 &14.1425  & &50000&14.25$\pm$0.13&14.11$\pm$0.12&14.16$\pm$0.12&14.17$\pm$0.13\\
				$S_0=110 $&0.0050 &14.1420 &14.1401  &14.0909$\pm 0.03$ &100000&14.03$\pm$0.09&14.18$\pm$0.09&14.10$\pm$0.09&14.06$\pm$0.09
				\\
				&0.0025 &14.1419 &14.1399  & &200000&14.05$\pm$0.06&14.04$\pm$0.06&14.07$\pm$0.06&14.13$\pm$0.06\\
				\hline
				& 0.02&22.2466 &22.2505  & &10000&22.38$\pm$0.32&22.84$\pm$0.33&22.46$\pm$0.32&22.15$\pm$0.32 \\
				&0.01 &22.2412 &22.2419  & &50000&22.35$\pm$0.15&22.27$\pm$0.14&22.24$\pm$0.14&22.28$\pm$0.14\\
				$S_0=120 $&0.0050 &22.2398 &22.2402  &22.1736$\pm  0.03$ &100000&22.12$\pm$0.10&22.27$\pm$0.10&22.19$\pm$0.10&22.17$\pm$0.10
				\\
				&0.0025 &22.2394 &22.2397  & &100000&22.12$\pm$0.10&22.27$\pm$0.10&22.19$\pm$0.10&22.17$\pm$0.10\\
				\hline
		\end{tabular}}
	}\quad\quad
	\subtable[]{
		\footnotesize
		{\begin{tabular} {@{}c|lrr|c|rrrrr@{}} \toprule $\rho_{Sr}=0.5$ & $\Delta y$ &HTFDa
				&HTFDb & B-AMC &$N_{\mathrm{MC}} $&HMCLSa  &HMCLSb &AMCLSa  &AMCLSb\\
				\hline
				& 0.02&1.3551 &1.3470  & &10000&1.18$\pm$0.09&1.29$\pm$0.10&1.12$\pm$0.09&0.80$\pm$0.08\\
				&0.01 &1.3576 &1.3566  & &50000&1.35$\pm$0.05&1.17$\pm$0.04&1.33$\pm$0.05&1.25$\pm$0.05\\
				$S_0=80$&0.0050 &1.3573 &1.3570  &1.3559$\pm0.01$ &100000&1.33$\pm$0.03&1.30$\pm$0.03&1.33$\pm$0.03&1.27$\pm$0.03 \\
				&0.0025 &1.3571 &1.3569  & &200000&1.35$\pm$0.02&1.31$\pm$0.02&1.38$\pm$0.02&1.34$\pm$0.02\\
				\hline
				& 0.01&3.7696 &3.7606  & &10000&3.72$\pm$0.17&3.78$\pm$0.17&3.82$\pm$0.18&3.72$\pm$0.17 \\
				&0.01 &3.7705 &3.7688  & &50000&3.86$\pm$0.08&3.71$\pm$0.08&3.80$\pm$0.08&3.81$\pm$0.08\\
				$S_0=90 $&0.0050 &3.7694 &3.7685  &3.7633$\pm0.02$ &100000&3.75$\pm$0.06&3.74$\pm$0.05&3.76$\pm$0.05&3.74$\pm$0.05\\
				&0.0025 &3.7694 &3.7686  & &200000&3.75$\pm$0.04&3.74$\pm$0.04&3.80$\pm$0.04&3.79$\pm$0.04\\
				\hline
				& 0.02&8.1285 &8.1249  & &10000&8.12$\pm$0.24&8.52$\pm$0.26&8.25$\pm$0.26&8.15$\pm$0.25\\
				&0.01 &8.1308 &8.1301  &  &50000&8.25$\pm$0.11&8.08$\pm$0.11&8.15$\pm$0.11&8.18$\pm$0.11\\
				$S_0=100 $&0.0050 &8.1311 &8.1308  &8.1122$\pm0.03$ &100000&8.07$\pm$0.08&8.16$\pm$0.08&8.11$\pm$0.08&8.10$\pm$0.08\\
				&0.0025 &8.1312 &8.1309  &  &200000&8.08$\pm$0.06&8.07$\pm$0.06&8.14$\pm$0.06&8.16$\pm$0.06\\
				\hline
				& 0.02&14.4455 &14.4468  & &10000&14.48$\pm$0.32&14.84$\pm$0.33&14.43$\pm$0.32&14.51$\pm$0.32\\
				&0.01 &14.4409 &14.4414  & &50000&14.60$\pm$0.15&14.40$\pm$0.14&14.45$\pm$0.14&14.47$\pm$0.14\\
				$S_0=110 $&0.0050 &14.4389 &14.4395  &14.3884$\pm0.03$ &100000&14.34$\pm$0.10&14.47$\pm$0.10&14.39$\pm$0.10&14.38$\pm$0.10 \\
				&0.0025 &14.4388 &14.4394  & &200000&14.35$\pm$0.07&14.37$\pm$0.07&14.38$\pm$0.07&14.48$\pm$0.07 \\
				\hline
				& 0.02&22.2859 &22.2893  & &10000&22.23$\pm$0.36&22.87$\pm$0.39&22.45$\pm$0.36&22.29$\pm$0.35\\
				&0.01 &22.2815 &22.2827  & &50000&22.50$\pm$0.17&22.29$\pm$0.16&22.27$\pm$0.16&22.28$\pm$0.16\\
				$S_0=120 $&0.0050 &22.2802 &22.2813  &22.2039$\pm0.04$ &100000&22.17$\pm$0.12&22.31$\pm$0.12&22.24$\pm$0.12&22.22$\pm$0.12\\
				&0.0025 &22.2798 &22.2808 &  &200000&22.17$\pm$0.08&22.17$\pm$0.08&22.17$\pm$0.08&22.32$\pm$0.08\\
				\hline
		\end{tabular}}
	}
	\caption{\em \small{Bates-Hull-White model. Prices of American call options. Test parameters: $K=100$, $T=0.5$,
			$\eta=0.05$, $r_0=0.03$, $\kappa_r=1$, $\sigma_r=0.2$, $V_0=0.04$, $\theta_V=0.04$, $\kappa_V=2$, $\sigma_V=0.4$,
			$\lambda=5$, $\gamma=0$, $\delta=0.1$, $\rho_{SV}=-0.5$,$\rho_{Sr}=-0.5, 0.5$.}}
	\label{tab7}
\end{table}

\begin{table} [ht] \centering
	\footnotesize
	{\begin{tabular} {@{}llrr|rccccc@{}} \toprule & $\Delta y$ &HTFDa &HTDFb &$N_{\mathrm{MC}}$&HMCa &HMCb &AMCa &AMCb\\
			\hline
			& 0.02  &2.77 &22.95 &10000 &0.13  &0.25  &0.36  &0.48  & \\
			& 0.01  &6.15  &48.17 &50000  &0.66  &1.35  &1.11  &2.48  &\\
			& 0.005  &12.12 &99.19  &100000 &1.37  &2.56  &1.82  &4.99  &\\
			& 0.0025  &27.61 &204.88 &200000  &2.56  &5.08  &3.70  &9.96  &\\
	\end{tabular}}
	\caption{\em \small{Bates-Hull-White model. Computational times (in seconds) for European call
			options in Table \ref{tab6} for $S_0=100$, $\rho_{Sr}=-0.5.$}}
	\label{tab81}
\end{table}
\begin{table} [ht] \centering
	\footnotesize
	{\begin{tabular} {@{}llrr|rccccc@{}} \toprule & $\Delta y$ &HTFDa &HTDFb &$N_{\mathrm{MC}}$&HMCLSa &HMCLSb &AMCLSa &AMCLSb\\
			\hline
			& 0.02  &2.77 &23.10 &10000 &0.28  &0.43  &0.40  &0.62  & \\
			& 0.01  &6.39  &48.65 &50000  &0.80  &1.79  &1.30  &2.72  &\\
			& 0.005  &12.50 &99.85  &100000 &1.91  &3.89  &3.02  &6.15  &\\
			& 0.0025  &27.92 &205.60 &200000  &4.03  &8.11  &5.20  &10.75  &\\
	\end{tabular}}
	\caption{\em \small{Bates-Hull-White model. Computational times (in seconds) for American call
			options in Table \ref{tab7} for $S_0=100$, $\rho_{Sr}=-0.5.$}}
	\label{tab82}
\end{table}

\clearpage

\section{Conclusions}

In this paper we have studied the numerical stability of  the hybrid tree/finite-difference method already introduced in \cite{bcz,bcz-hhw}. We have extended the method to the Bates model with a possible stochastic interest rate. We have also considered a Monte Carlo approach.
We have used our numerical schemes to evaluate European and American options. The results can be considered good and reliable, and the comparison with existing pricing methods has shown a good efficiency also in terms of computing time costs. 

%

\medskip

\noindent
\textbf{Acknowledgements.}
The authors wish to thank Andrea Molent for having implemented the
Alfonsi simulation scheme and the Monte Carlo Longstaff-Schwartz algorithms.


\end{document}